\documentclass[acmtog]{acmart}
\acmSubmissionID{1773}

\usepackage{booktabs} %

\usepackage{stfloats}
\usepackage{subcaption}
\usepackage{multirow}

\usepackage[ruled]{algorithm2e} %

\SetAlFnt{\small}
\SetAlCapFnt{\small}
\SetAlCapNameFnt{\small}
\SetAlCapHSkip{0pt}

\usepackage{bibunits}
\usepackage{float}

\setcopyright{cc}
\setcctype{by}
\acmJournal{TOG}
\acmYear{2026} \acmVolume{45} \acmNumber{4} \acmArticle{}
\acmMonth{7} \acmDOI{10.1145/3811402}

\begin{document}
\title{MUSIC: Learning \textbf{Mus}cle-Dr\textbf{i}ven Dexterous Hand \textbf{C}ontrol}

\author{Pei Xu}
\authornote{Both authors contributed equally to this research.}
\orcid{0000-0001-7851-3971}
\affiliation{%
 \institution{Stanford University}
 \country{USA}}
\email{peixu@stanford.edu}
\author{Yufei Ye}
\authornotemark[1]
\orcid{0000-0001-8767-0848}
\affiliation{%
 \institution{Stanford University}
 \country{USA}}
\author{Shuchun Sun}
\orcid{0000-0002-7985-2745}
\affiliation{%
 \institution{Clemson University}
 \country{USA}}
\author{Yu Ding}
\orcid{0009-0008-4759-5677}
\affiliation{%
 \institution{Stanford University}
 \country{USA}}
\author{Elizabeth Schumann}
\orcid{0009-0007-1331-2707}
\affiliation{%
 \institution{Stanford University}
 \country{USA}}
\author{C. Karen Liu}
\orcid{0000-0001-5926-0905}
\affiliation{%
 \institution{Stanford University}
 \country{USA}}

\begin{abstract}
We present a data-driven approach for physics-based, muscle-driven dexterous control that enables musculoskeletal hands to perform precise piano playing for novel pieces of music outside the reference dataset. Our approach combines high-frequency muscle-level control with low-frequency latent-space coordination in a hierarchical architecture. At the low level, general single-hand policies are trained via reinforcement learning to generate dynamic muscle-tendon activations while tracking trajectories from a large reference motion dataset. 
The resulting tracking policies are then distilled into variational autoencoder (VAE) models, yielding smooth and structured latent spaces that abstract away low-level muscle dynamics.
For the high level, we train piece-specific policies to operate in this latent space, coordinating bimanual motions based on specific goals, denoted by note events extracted from given musical scores, to synthesize performances beyond the reference data. High-level control is formulated as a decentralized multi-agent reinforcement learning problem combined with adversarial learning for motion imitation. 
In addition, we present an enhanced musculoskeletal hand model that supports fine control of fingers for accurate low-level motion tracking and diverse high-level motion synthesis.
We evaluate the control pipeline of our approach on a diverse piano repertoire spanning multiple musical styles and technical demands. Results demonstrate that our approach can synthesize coordinated bimanual motions with accurate key presses, and achieve the state-of-the-art performance of piano playing in physics-based dexterous control, while generalizing to sheet music that is not presented in the reference dataset.
We also show that our musculoskeletal hand model demonstrates superior biomechanical stability and tracking precision compared to the existing model, and validate that our musculoskeletal hand model and muscle-driven controller can generate physiologically plausible activation patterns that align with human electromyography (EMG) recordings when subjects perform multiple tasks.

\end{abstract}

\ccsdesc[500]{Computing methodologies~Animation}
\ccsdesc[300]{Computing methodologies~Physical simulation}
\ccsdesc[300]{Computing methodologies~Reinforcement learning}

\keywords{character animation, physics-based control, motion synthesis, hierarchical reinforcement learning}

\begin{teaserfigure}
\centering
    \includegraphics[width=\linewidth]{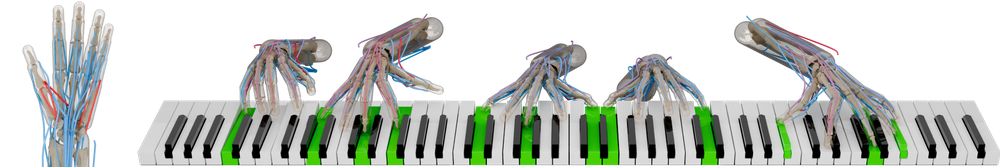}
    \caption{Our musculoskeletal hand model (left) and diverse hand poses (right) produced by our muscle-driven motion synthesis models during piano playing. The muscle-tendon units are visualized by blue lines, with the activated ones highlighted in red. The semi-transparent shell indicates the outer skin (collision geometry) of the hand model.}
 \label{fig:teaser}
\end{teaserfigure}

\maketitle

\section{Introduction}

Physically synthesizing human motion has a wide range of applications in character animation, embodied AI, AR/VR, robotics, and biomechanics. 
Within these domains, musculoskeletal models represent a distinctive and valuable research direction. 
Unlike traditional joint-actuated models, muscle-driven simulation provides a more faithful modeling of biological human musculoskeletal systems bounded by physiological constraints. 
Besides improving the realism of character animation,
effective control of muscle-driven models offers unique opportunities to analyze muscle fatigue, quantify injury risk, and optimize motion performance, contributing to biomechanics and biomedical engineering and providing insights and control solutions to biorobotics.

This work advances the control and modeling of muscle-driven dexterous manipulation through the lens of elite piano performance. Piano playing exemplifies peak human dexterity, requiring extreme precision, rapid bimanual coordination, and sophisticated motor control, all while operating within strict physiological constraints imposed by muscle–tendon dynamics. 
Prior research has explored synthesizing piano performance in digital agents. 
While some approaches~\cite{liu2025separate,gan2024pianomotion10m} focus on generating hand motions in kinematic space, another line of research~\cite{zakka2023robopianist,wang2024furelise} synthesizes piano playing through physical simulation to generate physically plausible motions.
However, existing state-of-the-art systems rely exclusively on control over the joint space, ignoring the underlying muscle dynamics and frequently producing biologically unrealistic joint poses.

In this work, we take a first step toward muscle-driven hand motion synthesis for elite-level piano performance. Given a musical score denoted by note events, %
our method directly controls bimanual, muscle-actuated hands to execute the required notes with high precision. This task is particularly challenging because muscle-driven systems 
are naturally over-actuated and 
exhibit complex nonlinear dynamics
with unidirectional actuatable drivers (musculotendon units).
While reinforcement learning is a common way to solve physics-based control,
those characteristics result in an action space, which is not friendly to reinforcement learning, 
and also demand a high-frequency control system to provide precise coordination of hand and finger poses.

To address these challenges, we propose a two-level hierarchical control architecture that separates general muscle coordination from music-specific decision making.
Specifically, our framework consists of:
(1) a low-level controller that directly actuates musculotendon units by outputting activations at high frequency, trained to imitate poses in a large-scale piano performance dataset; and
(2) a high-level policy, trained per musical piece, that translates musical score input into target behaviors while remaining agnostic to the underlying muscle control.
The high-level policy operates at a lower frequency without needing to know the low-level states of the muscle-tendon units, largely increasing training efficiency.
Instead of coupling the two controllers via explicit physical quantities such as joint targets or torques, we introduce an implicit latent-space interface through variational autoencoder (VAE) models.
This design mitigates out-of-distribution exploration in the high-dimensional physical state space and promotes stable, physically plausible motion synthesis.

In parallel, we enhance the musculoskeletal hand model to support the high-precision demands of piano playing. While current models may be sufficient for common grasping and in-hand manipulation tasks, %
we find that they cannot meet %
the dexterity demands of piano performance. In particular, we identify several critical muscles missing from existing models, leading to inadequate abduction-adduction and flexion strength in fingers. %
Guided by biomechanics literature, we augment the hand model with additional muscles 
at the lateral sides of the palm, improving  fine-grained control of the thumb and pinky fingers and the overall stability of hand posing.

By integrating our hierarchical control algorithm with the enhanced musculoskeletal model, we present the first system capable of synthesizing hand motions for piano performance that are both physically and biomechanically plausible. Our method demonstrates high precision of dexterous control, achieving F1 scores above 0.9 on a challenging testing repertoire.
To validate our design, we perform additional experiments to evaluate the hand musculoskeletal model, and conduct extensive sensitivity analysis to verify the latent-space interface of the hierarchical control.

\section{Related Work}

Early research into animating musical performance primarily relied on heuristics and kinematic optimization to solve for feasible postures of instrument playing~\cite{zhu2013system,elkoura2003handrix}.
In recent years, data-driven approaches using generative models~\cite{liu2025separate,gan2024pianomotion10m,Liu2020BodyMG, shlizerman2018audio,li2018skeleton, kao2020temporally,chen2021guzheng,qiu2025elgar,canales2025real,kyriakou2025drums} have drawn lots of attention.
However, due to the lack of physical constraints, those methods working on kinematics motion generation often suffer from the problem of artifacts.
Our work follows the literature of physics-based dexterous manipulation with a focus on muscle-driven control to generate high-fidelity and physically plausible motions.

\subsection{Dexterous Control for Musical Performance}

Prior work on dexterous control has primarily focused on two broad categories: object grasping and relocation~\cite{xie2023hierarchical,zhao2013robust,liu2009dextrous}
and in-hand manipulation~\cite{andrychowicz2020learning,zhang2021manipnet,yang2022learning}.
Musical performance introduces a distinct and more demanding form of dexterous control, where fine-grained finger articulation must be tightly synchronized with precise timing, continuous contact, and coordinated bimanual motion.
Previous studies have explored the tasks of playing guitar~\cite{luo2024learning,guitar}, drum~\cite{shahid2025robot}, and piano~\cite{xu2022towards,zakka2023robopianist,wang2024furelise}.
Due to the complexity of the task, for piano playing,~\citet{xu2022towards} only considers one hand playing on a simplified piano;~\citet{zakka2023robopianist} needs human-annotated fingering to guide policy learning; and~\citet{wang2024furelise} relies on a diffusion model to generate reference motions to facilitate policy training.
Our approach can directly synthesize bimanual piano-playing motions from a large-scale reference dataset of more than 10 hours, without requiring fingering annotations or auxiliary motion generators.

\subsection{Physics-based Muscle-driven Control}
For muscle-driven control, early work largely relied on nonlinear optimization, analytical models, and hand-crafted heuristics to perform control~\cite{ackermann2010optimality,anderson2001dynamic,geijtenbeek2013flexible,ong2019predicting,song2015neural,song2018predictive,wang2012optimizing,waterval2021validation,dembia2020opensim,falisse2019rapid,geyer2010muscle,lee2014locomotion,tsang2005helping,jiang2019synthesis,sachdeva2015biomechanical,sueda2008musculotendon}. %
Recently, reinforcement learning has emerged as a powerful alternative, enabling impressive progress in muscle-driven control by learning activation strategies through interaction with physics simulators.
However, the over-actuation nature of musculoskeletal systems imposes additional challenges.
Several approaches~\citep {zuo2024self,he2024dynsyn,berg2024sar} exploit muscle synergies to reduce the action space.
While effective for training stabilization, such approaches constrain multiple musculotendon units to fixed activation patterns, thereby limiting expressiveness and fine-grained control. 
Another line of work converts muscle activations into joint-space control via intermediate joint-level policies~\cite{lee2019scalable,park2022generative}, or employs a PD controller over tendon length to generate activations ~\cite{si2014realistic,feng2023musclevae}.
Those methods explicitly leverage muscle dynamics equations during training or control. 
In contrast, our approach treats muscle dynamics as a black box and learns control policies purely through interaction with the simulator, making the learning process independent of the simulator’s internal muscle model and avoiding the need to solve muscle inverse dynamics.
Moreover, while the prior literature focuses on motion tracking or locomotion, our method enables the synthesis of complex, high-precision bimanual motions with strict spatial and temporal requirements, as exemplified by piano performance.

\subsection{Hierarchical Control for Physics-Based Animation}
Hierarchical control has been explored in physics-based character animation to address high-dimensional control and long-horizon behaviors. Methods such as MotionVAE~\cite{ling2020character}, PULSE~\cite{luouniversal}, MaskedMimic~\cite{tessler2024maskedmimic}, and PhysicsVAE~\cite{won2022physics} learn latent spaces for high-level, task-directed control while relying on low-level controllers for motion execution, and FreeMusco~\cite{kim2025freemusco} extends this idea to muscle-driven characters. Other works, including~\citet{won2021control} and~\citet{zhu2023neural}, employ hierarchical structures for multi-agent interaction, while MCP~\cite{peng2019mcp} and Learn-to-Ball~\cite{xu2025learning} focus on policy composition.
In contrast, our approach targets muscle-driven dexterous manipulation with strict precision requirements. 
We introduce a dual-frequency hierarchical design with latent-space distillation to decouple high-level motion synthesis from low-level dynamic control of muscles. 
The low-level controller is treated as a general-purpose musculotendon controller rather than a generator of fixed motion patterns, and we incorporate imitation learning during high-level training. This design enables stable control and scales to coordinated, contact-rich bimanual tasks.

\section{Overview}

\begin{figure*}
    \centering
    \includegraphics[width=.9\linewidth]{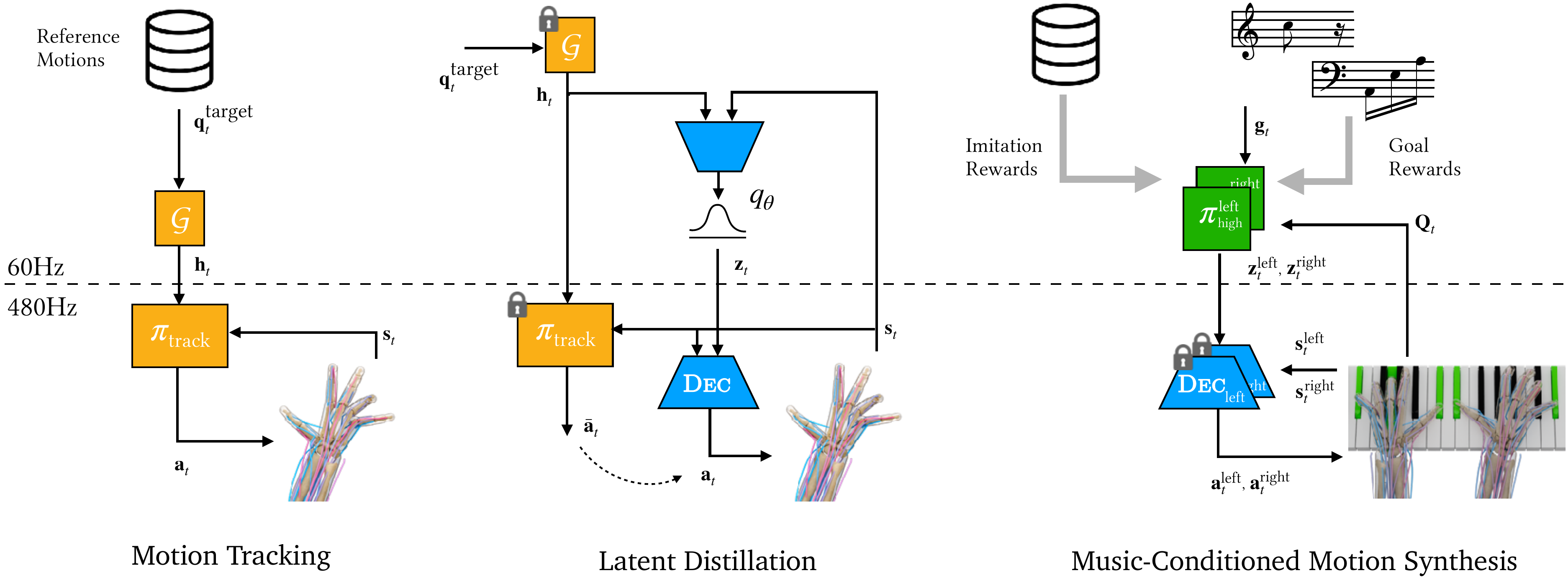}
    \caption{System Overview.
    The whole system of our framework is trained in three stages.
    First, we learn a single-hand tracking policy $\pi_{\text{track}}$ that outputs high-frequency muscle activations for direct muscle-driven control (Sec.~\ref{sec:track}).
    Second, we perform on-policy distillation to obtain a VAE decoder as a low-level servo for muscle control, while taking as input the well-structured latent action $\mathbf{z}_t$ (Sec.~\ref{sec:vae}). 
    Lastly, we train a piece-specific high-level controller over the latent to synthesize motions for piano playing (Sec.~\ref{sec:high}).
    While the first and third stages are conducted through reinforcement learning, the second stage is performed through supervised training.
    Specifically, for high-level controller training in the third stage, we adopt a decentralized multi-agent setting to treat two hands as individual agents associated with independent policies $\pi_\text{high}^\text{left}$ and $\pi_\text{high}^\text{right}$, while taking a shared kinematic observation of two hands for coordinated control.
    }\label{fig:overview}
\end{figure*}

Given a musical score, our framework generates muscle activations to actuate bimanual musculoskeletal hand models.
As illustrated in Fig.~\ref{fig:overview}, the system is organized as a hierarchical architecture with components operating at different temporal resolutions, and is trained in three stages.
In the first stage, for each hand, we train a general-purpose tracking policy that imitates poses from a large-scale piano performance dataset. The policy takes as input target poses and the current proprioceptive state of the hand, and outputs high-frequency muscle activations that directly actuate the simulated model.
In the second stage, the tracking policies are distilled into Variational Autoencoders (VAEs), each of which maps the high-dimensional target state space for a single hand into a compact latent representation updated at a lower frequency.
In the third stage, we train a decentralized high-level controller that takes as input kinematic hand states and musical goals represented as sequences of note events, and outputs latent actions at a lower frequency. This latent interface enables the high-level controller to coordinate bimanual motion while remaining decoupled from low-level muscle dynamics.

In addition, we introduce a more functional and biomechanically realistic musculoskeletal hand model~(Sec.~\ref{sec:hand_model}).
Based on prior work~\cite{caggiano2022myosuite}, we augment the model with additional musculotendon units located along the lateral regions of the palm. These additions improve abduction–adduction capability and fine control of the thumb and pinky, leading to more stable and accurate hand postures during dexterous manipulation.

\section{General Muscle-Driven Tracking}
\label{sec:track}

In the first stage, we train a general muscle-driven policy to track any given target trajectory $\mathbf{q}_t^\text{target}$ indicating desired hand poses in $H$ frames. The policy, denoted as $\pi_{\text{track}}(\mathbf{a}_t | \mathbf{s}_t, \mathbf{h}_t)$, is jointly learned with a network $\mathcal{G}: \mathbf{q}_t^\text{target}\mapsto \mathbf{h}_t$, which encodes the target trajectory into a pose embedding $\mathbf{h}_t$. The policy maps the current proprioception~$\mathbf{s}_t$ and the encoded target pose $\mathbf{h}_t$ to muscle activations $\mathbf{a}_t$ at 480Hz, which is at the same frequency as the physics simulation. %

To ensure stable state initialization and faster convergence, the tracking policy is trained purely for motion tracking in an environment without the piano. We train the policy on FürElise \cite{wang2024furelise}, a large-scale dataset featuring elite piano performances, comprising approximately 10 hours of hand motions. The dataset follows a long-tail distribution with specialized techniques like glissando and uncommon palm-up gestures, while most trajectories consist of various finger lifting and pressing motions
for piano playing.

\vspace{\baselineskip}\noindent\textbf{Dual-Frequency Tracking Policy.} We design the tracking policy to operate at two different frequencies. The policy outputs muscle activations at a high frequency of 480Hz, while the target poses over $H$ frames are updated at a lower frequency of 60Hz, resulting in $\mathbf{q}_t^\text{target}$ containing an $H$-frame trajectory with a sampling interval of $n=8$ frames.
This dual-frequency design is critical for meeting the high-frequency control demands of the hand musculoskeletal system, while allowing the subsequent high-level controller to operate at a lower frequency for improved training and inference efficiency.

\vspace{\baselineskip}\noindent\textbf{Observation Space.} 
In addition to the target pose embedding $\mathbf{h}_t$, the network $\pi_\text{track}$ receives the proprioceptive state $\mathbf{s}_t$ as input. This state comprises link positions, orientations, and linear and angular velocities, as well as each musculotendon unit's current length, velocity, and activation. 
While tendon length and velocity are needed to evaluate the Force–Length–Velocity (FLV) relationship~\cite{uchida2021biomechanics},
the current activation is also required, 
as the resulting activation depends on both the existing activations and newly applied inputs through activation dynamics governed by time constants~\cite{millard2013flexing}.

\vspace{\baselineskip}\noindent\textbf{Reward Definition.}
The tracking policy is optimized to achieve precise motion imitation while minimizing energy expenditure.
To that end, we define a tracking reward regularized to generate sparse muscle activations:
\begin{equation}\label{eq:tracking_reward}
    r_t = 0.9 r_\text{tracking} + 0.1 r_\text{act},
\end{equation}
where
\begin{equation}\begin{split}
    r_\text{tracking} &= 0.5 r_\text{pos} + 0.5 r_\text{orient} \\
    r_\text{pos} &= 0.7\exp(-50 e_{p})+ 0.3\exp(-3 e_{p}) \\
    r_\text{orient} &= \exp(-3 e_{o}) \\
    r_\text{act} &= \exp(-||\mathbf{a}_t||_4^4/\dim \mathbf{a}_t).
\end{split}
\end{equation}
The position error $e_{p} = \sum_i \kappa_i ||\mathbf{p}_i^\text{target} - \mathbf{p}_i||$ measures the distance of each link to its immediate target,
and the orientation error $e_{o} = \sum_i \omega_i (\mathbf{r}_i^\text{target} \ominus \mathbf{r}_i)^2$ measures the squared angle difference between two orientations, with weights $\kappa_i$ and $\omega_i$.
The two terms in $r_\text{pos}$ allow the reward to capture both long-distance errors and millimeter-scale precision, accounting for the large spatial range of a piano keyboard (approximately $122$cm) relative to the small width of individual keys ($1.7$–$2.35$cm).
We omit the subscript $t$ in the equations for clarity.

During simulation, the action $\mathbf{a}_t$ is clipped to the range $[0,1]$, consistent with the standard definition of muscle activations. 
However, its raw, \emph{unclipped} value is used during reward computation in $r_\text{act}$.
The fourth-order norm in $r_\text{act}$ penalizes large activations, while remaining tolerant to small ones.
This formulation discourages excessive muscle exertion without forcing the policy to trade off tracking accuracy when approaching the target pose. 
While the network is allowed to provide sparse control signals by outputting negative values,
we penalize large negative values in $\mathbf{a}_t$, %
preventing the learning process from drifting toward biased negative regions.
Overall, our tracking policy can produce sparse, human-like activation patterns while keeping high precision of motion tracking.

\begin{figure}
    \centering
    \includegraphics[width=\linewidth]{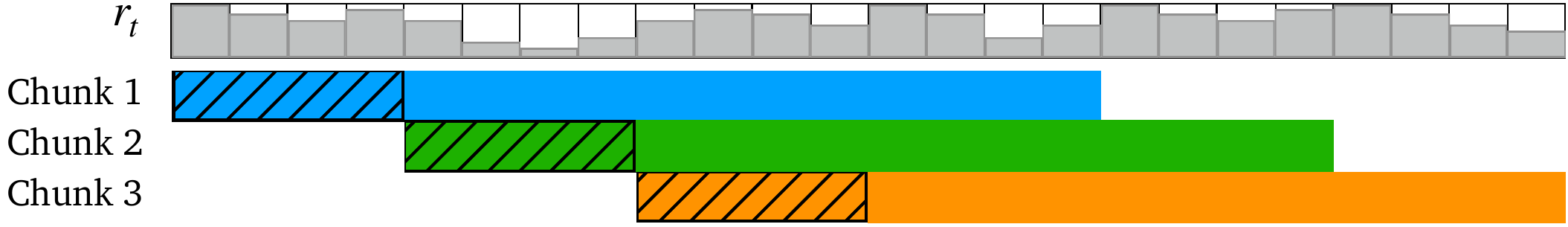}
    \caption{A demonstration trajectory is divided into three overlapping chunks. The shadowed regions indicate the sampling regions of each chunk. The top bars exhibit the policy's tracking performance per frame. Our MCMC sampling scheme lets the frames near the sampling region of each chunk contribute more to the performance estimation, and thus help localize the chunk that starts with badly performed frames. For example, here, Chunk 2 will have a lower running performance estimation than Chunk 1, because the badly tracked frames are in the sampling region of Chunk 2, though also covered by Chunk 1.}
    \label{fig:adp_sample}
\end{figure}

\vspace{\baselineskip}\noindent\textbf{Adaptive Sampling.} 
To effectively learn from the diverse reference data, we adopt an adaptive sampling strategy based on a power-scaled MCMC scheme.
As shown in Fig.~\ref{fig:adp_sample}, we first partition each reference trajectory (thousands of frames) into overlapping chunks which are roughly equally long with a maximal length of $C$ frames. During training, a chunk is sampled, and the agent is initialized at a state with target poses drawn uniformly from the chunk's \textit{non-overlapping} region. 
The training episode will end
early if it reaches the end of that chunk, or terminate unsuccessfully
if the tracking error becomes too large, resulting in an episode with
$L$ frames long, where $L \leq C$.
We track the policy's running performance $\bar{r}_i^\prime$ through an MCMC approach.
For a given chunk $i$,
we estimate the policy's performance online by calculating the discounted cumulative reward $\bar{r}_i^\prime = \frac{C}{L}\sum_{t=0}^{L} \zeta^t r_t$, where $\zeta$ is a discount factor.
If the episode terminates unsuccessfully, we assign $r_t = 0$ for all the remaining frames.
The recorded performance $\bar{r}_i$ is updated through a moving average at the end of each episode, i.e., $\bar{r}_i \leftarrow (1-\alpha) \bar{r}_i + \alpha \bar{r}_i^\prime$.

The sampling weight of chunk $i$ is decided by
\begin{equation}\label{eq:adapt_sampling}
    w_i = \frac{1}{Z} \left[\left( \frac{r_\text{max}-r_\text{min}+\epsilon}{\bar{r}_i-r_\text{min}+\epsilon}\right)^\eta - 1 + \varepsilon\right]
\end{equation}
where $Z$ is a normalization factor, and $\eta \geq 1$ controls the sharpness of the distribution, amplifying the probability for chunks with low $\bar{r}_i$. The terms $r_\text{max}$ and $r_\text{min}$ represent the theoretical reward bounds, while $\epsilon$ and $\varepsilon$ are small constants added for numerical stability. All $\bar{r}_i$
values are initialized using a small positive constant before training.
This sampling strategy ensures that the policy explicitly practices the specific chunks starting with badly performed frames, as the discounted reward heavily weighs these initial frames.

\section{Latent Distillation}
\label{sec:vae}

While the target pose embedding $\mathbf{h}$ could be directly leveraged to drive the tracking policy, our experiments indicate that the latent defined in the unconstrained space of $\mathbf{h}$ is still too large to provide densely distributed valid control points for efficient exploration by a high-level controller.
Therefore, we further distill this embedding into a smooth manifold $\mathbf{z}$ to facilitate high-level controller training.

We distill the tracking policy through a VAE model, where the decoder mimics the tracking policy $\pi_\text{track}$ by mapping a slow latent $\mathbf{z}_t$ to high-frequency muscle activations.
Following $\beta$-VAE~\cite{higgins2017beta}, we perform distillation by minimizing the loss function
\begin{equation}\label{eq:vae}
    \mathcal{L} = \mathbb{E}_t\Big[%
    \sum_{\delta=0}^{n-1} ||\textsc{Dec}(\mathbf{s}_{t+\delta}, \mathbf{z}_t) - \mathbf{\bar{a}}_{t+\delta} ||^2  
     + \beta D_{KL}(q_\theta(\mathbf{z}_t|\mathbf{s}_t, \mathbf{h}_t)||p)\Big] 
\end{equation}
where $\mathbf{\bar{a}}_{t+\delta} = \textsc{Clip}(\mathbf{a}_{t+\delta}, 0, 1)$ with $\mathbf{a}_{t+\delta} \sim \pi_\text{track}(\cdot | \mathbf{s}_{t+\delta}, \mathbf{h}_t)$, which is the target action provided by the tracking policy at the high frequency,
$\mathbf{z}_t \sim q_\theta(\cdot|\mathbf{s}_t, \mathbf{h}_t)$ is the latent code sampled every $n$ frame from the encoder posterior $q_\theta$, modeled as multi-variate Gaussian,
$\textsc{Dec}$ denotes the decoder network, and $p$ is the prior.
We use a standard normal prior $p:=\mathcal{N}(\mathbf{0}, \mathbf{I})$,
which performs slightly better than a conditional prior $p(\cdot|\mathbf{s}_t)$ during our experiments.

While the decoder is trained as a deterministic model to imitate the behavior of $\pi_\text{track}$, we adopt a hybrid on-policy distillation strategy to improve robustness. While the target pose embedding $\mathbf{h}_t$ is always sampled from the reference motion data through $\mathcal{G}(\mathbf{q}_t^\text{target})$, the simulation state $\mathbf{s}_t$ is dynamically unrolled using activations provided by the VAE decoder and those from $\pi_\textbf{track}$ in a mixture way.
This unrolling exposes the VAE to its own execution errors, encouraging recovery behaviors and improving control robustness, while mitigating excessive state drift by applying activations from $\pi_\textbf{track}$ with a $20\%$ probability.
We refer to the supplementary materials for additional details.

Although we do not explicitly regularize muscle coactivation during distillation, the hierarchical architecture introduces a low-dimensional latent representation that implicitly regularizes control and encourages coordinated activation patterns.

\section{Music-Conditioned Motion Synthesis}\label{sec:high}
With the low-level muscle-driven controller trained for each hand through latent distillation, %
we %
learn a piece-specific high-level controller to coordinate bimanual motion generation for piano performance. 
Unlike the low-level controllers, which are trained in isolation for two hands, 
we train 
the high-level controller with two hands in
a shared environment that includes a piano.
While the low-level controllers were trained purely on motion tracking without a piano, they generalize effectively to this contact-rich environment, successfully executing the physical interactions required for pressing keys.
Instead of training a %
centralized controller to output actions %
for both hands, we optimize separate policies for each hand in a decentralized manner. %
Specifically, for high-level control, we employ two policies corresponding to the left and right hands, $\pi_\text{high}^\text{left}$ and $\pi_\text{high}^\text{right}$, respectively, each paired with its own low-level controller, while sharing a common kinematic observation to enable coordinated bimanual control.
Under this decentralized design, each hand acts as an independent agent that greedily maximizes its own reward while sharing kinematic pose information within the observation space.
This design significantly reduces the action-space complexity, %
leading to efficient policy training. %

\begin{figure}
    \centering
    \includegraphics[width=\linewidth]{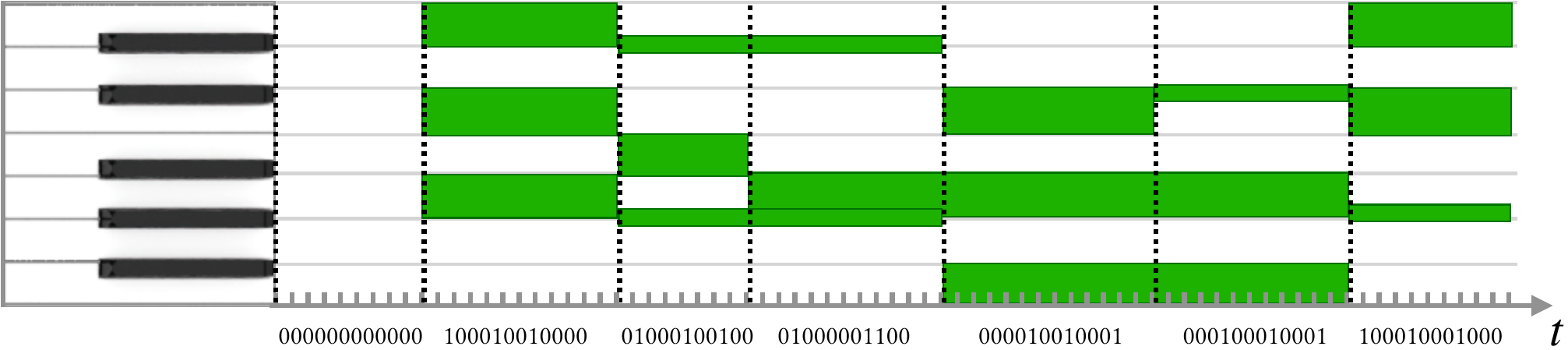}
    \caption{Demonstration of our music goal representation $\mathbf{g}_t$. The green boxes indicate the upcoming target keys. Instead of showing the target keys in a fixed number of frames, we use the next $N$ target key-pressing patterns as the goal input to the policy network to increase the policy's observation horizon.
    The dashed lines divide the target into different patterns. The annotation under each pattern denotes its encoded representation, while $\mathbf{g}_t$ represents the full keyboard with 88 keys, augmented with an additional $+/-$ sign to indicate hand assignment.}
    \label{fig:midi_goal}
\end{figure}

\vspace{\baselineskip}\noindent\textbf{Observation Space.} 
Each single-hand policy takes the same input of the note events as the goal $\mathbf{g}_t$,
alongside the kinematics of both hands $\mathbf{Q}_t$ to coordinate bimanual motion and prevent collisions, i.e., $\pi^h_{\text{high}}(\mathbf{z}_t^{h} | \mathbf{Q}_t, \mathbf{g}_t)$ where $h \in \{\text{left}, \text{right}\}$. The kinematics $\mathbf{Q}_t$ comprise a two-frame historical observation, including the position, orientation and velocities of each link of two hands. %
The goal $\mathbf{g}_t \in \mathbb{R}^{N \times (88+1)}$ is represented as an indicator vector of the next $N$ target key-pressing patterns (cf. Fig.~\ref{fig:midi_goal}), concatenated with a timer variable indicating the duration of each pattern \cite{guitar,wang2024furelise}. 
To distinguish target hands, we use a ternary representation instead of the binary representation used in prior work.
The goal based on note events provides a significantly longer horizon than 
a fixed-timestep target goal,
allowing the high-level controller to anticipate and plan for upcoming notes more effectively.

\vspace{\baselineskip}\noindent\textbf{Reward Definition.}
To ensure natural movement, we use imitation-based reinforcement learning while incorporating a GAN-like architecture~\cite{iccgan} to perform motion imitation, and adopt a multi-objective learning framework~\cite{xu2023composite} to balance motion imitation and goal-directed control.
In the decentralized multi-agent setting, rewards must be computed separately for each hand, which requires knowledge of hand-level key assignments.
In our implementation, we use a clef-based heuristic to determine hand assignment from a given musical score, following the common convention that the bass clef is assigned to the left hand and the treble clef to the right hand. 
The task reward for each hand consists of positive terms $r_k^+$ that encourage correct key presses and negative terms $r_k^-$ that penalize unintended presses: 
\begin{equation}\label{eq:goal_reward}
    r_{t,h} = \frac{1}{|\mathcal{K}_h|}\sum_{k\in\mathcal{K}_h} r_k^+ - 0.2 \sum_{k\notin\mathcal{K}} r_k^-
\end{equation}
where $\mathcal{K}_h$ represents the set of target keys for hand $h$.

For each target key $k \in \mathcal{K}_h$, we minimize the distance $\|\mathbf{d}_{k,i}\|$ between the key and the nearest finger $i$ while taking into account finger occupancy and spans, and encourage key depression: 
\begin{equation}\label{eq:r_k_pos}\begin{split}
    r_k^+ &= 0.6 r_k^\text{dist} + 0.4  r_k^\text{press} \\
    r_k^\text{dist} &= 0.8\exp(-500||\mathbf{d}_{k,i}||^2) + 0.2\exp(-5||\mathbf{d}_{k,i}||) \\
    r_k^\text{press} & = (d_k / D_k)^3 \mathbb{I}_{k,i}
\end{split}
\end{equation}
where $d_k$ is the current rotation angle of key $k$, and $D_k$ is its maximum rotation range. Specially, when $\mathcal{K}_h = \emptyset$, the positive reward is set to $1$. 
We consider an effective key pressing (generating sound) if $d_k / D_k > \tau$. 
Specifically, to discourage intermittent pressing,
we penalize temporary key release and set $r_k^\text{press} = 0$ if $d_k/D_k > \tau$ at timestep $t-1$ but $d_k/D_k \leq \tau$ at timestep $t$. We set $\tau = 0.9$ in our implementation and assume a maximal depth displacement of 1cm when a key is fully pressed down~\cite{van1995modeling}.
Notably, our approach can also support predefined finger-level key assignment when desired by replacing the nearest-finger heuristic with specified target fingers during reward computation, and by explicitly encoding the assigned finger index for each target key in the goal vector $\mathbf{g}_t$.

For non-target keys, we penalize the normalized depression $r_k^- = (d_k / D_k)^6$. The higher power encourages rapid release and penalizes deep presses while tolerating light resting contact. The negative terms are computed over all non-target keys across two hands, i.e. $\mathcal{K} = \cup_h \mathcal{K}_h$. We adopt this holistic penalty, since attributing incorrect presses to individual hands is non-trivial. 

To ensure natural movement, we additionally incorporate a GAN-like architecture~\cite{iccgan} to perform imitation learning from the reference motion data, and adopt a multi-objective learning framework~\cite{xu2023composite} to balance multiple imitation and goal-directed objectives.
We refer to the supplementary materials for the details of high-level controller training.

\vspace{\baselineskip}\noindent\textbf{Adaptive Sampling.} 
To improve the training efficiency, we adopt the adaptive sampling strategy elaborated in Sec.~\ref{sec:track}, but at a note level.
We estimate the high-level controller's performance by
\begin{equation}
    \bar{r}_i = \frac{C}{L}\sum_{t=0}^L \zeta^t \min \{\textsc{Recall}_t^\text{left}, \textsc{Recall}_t^\text{right}\}.
\end{equation}
Here, we utilize the recall score as the driver for
adaptive sampling, as the policy generally achieves high precision easily.
In each training episode, we use weights derived from Eq.~\ref{eq:adapt_sampling} to sample a starting note from the given musical score.
Each episode is $C$-frame long, but terminates early, resulting in an $L$-frame episode, if it reaches the end of the music score.
This sampling strategy focuses the learning process on difficult passages where the keys are often missed, ensuring more robust performance across the entire piece.

\begin{figure}
    \centering
    \includegraphics[width=0.53\linewidth]{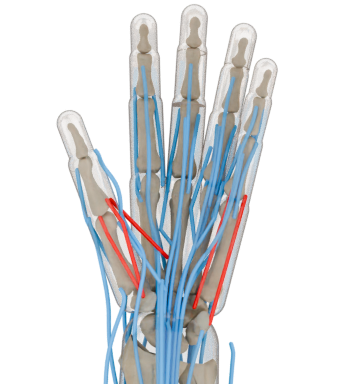}
    \includegraphics[width=0.45\linewidth]{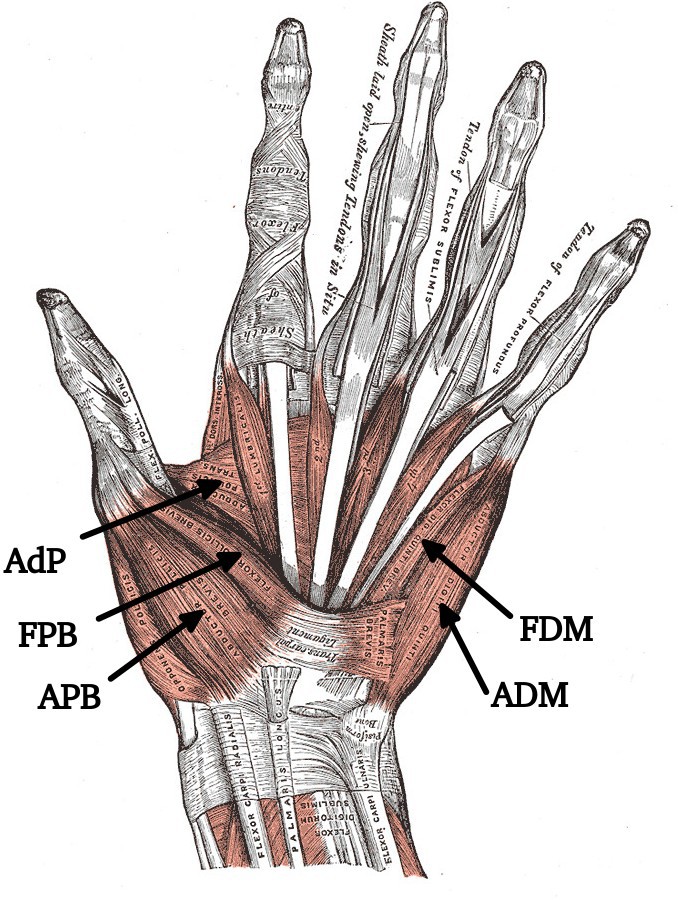}
    \caption{Illustration of our enhanced musculoskeletal hand model (left hand) with the added musculotendon units highlighted in red. The right subfigure from \textit{Gray's Anatomy} visualizes the volumes and locations of the muscles.}
    \label{fig:hand_anatomy}
\end{figure}

\section{Musculoskeletal Hand Model}
\label{sec:hand_model}

Our hand model is adapted from MyoHand~\cite{caggiano2022myosuite}, which provides anatomically grounded muscle–tendon structures and joint configurations based on previous work in biological modeling~\cite{mcfarland2019spatial,lee2015finger}. 
We keep
the forearm in the model while removing the rest skeletal bodies, as many muscles responsible for finger and wrist motion originate there.
A free root is introduced at the elbow position with 6 DoFs actuated through direct torque and force control. %
In addition, we augment the model with five tendon units, including FPB, AdP, APB, FDM, and ADM, located along the lateral regions of the palm, as highlighted in Fig.~\ref{fig:hand_anatomy}. 
The resulting model comprises 44 actuatable musculotendon units, 23 movable joints, and 34 rigid body links, yielding a total of 50 controllable DoFs per hand (44 muscle activations and 6 root DoFs).
The left-hand model is obtained by mirroring the right hand.
Additional details regarding anatomical modifications and control implications of our hand models are provided in the supplementary materials. 

We rely on MuJoCo to simulate muscle–tendon dynamics, where muscles and tendons are coupled as musculotendon units during simulation rather than explicitly modeling elastic tendons as in OpenSim~\cite{delp2007opensim}.
We also rely on MuJoCo’s implementation for activation dynamics~\cite{millard2013flexing}. The model uses default activation and deactivation time constants of $10$ms and $40$ms, governing how quickly muscle activation rises and falls in response to controller inputs rather than assuming instantaneous muscle response.

\section{Experiments}

We use MuJoCo~\cite{todorov2012mujoco} as the physics simulator, which provides built-in support for musculotendon simulation and actuation. All simulations are performed using the XLA-based MuJoCo implementation (MJX) with GPU computation support through JAX. The neural networks are implemented using PyTorch.
Additional analyses about the key design choices of our approach are provided in the supplementary materials.

\subsection{Muscle-Driven Motion Tracking}
\begin{figure*}
\begin{minipage}{\columnwidth}
    \centering
    \includegraphics[width=\linewidth]{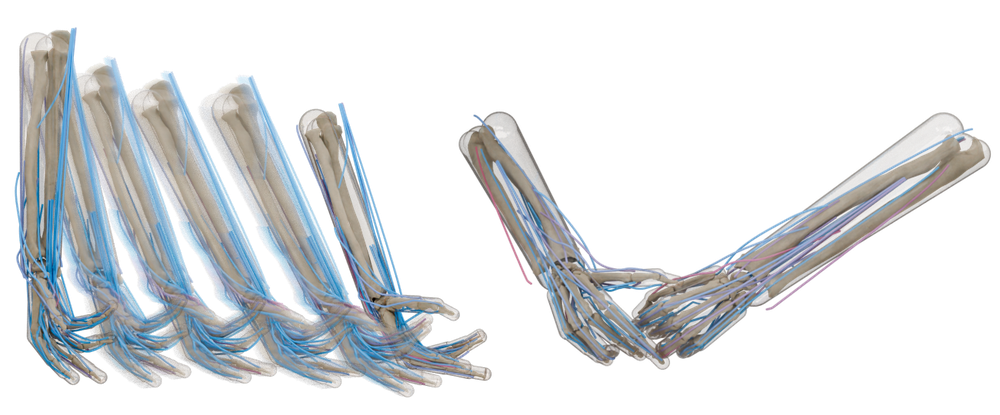}
    \includegraphics[width=\linewidth]{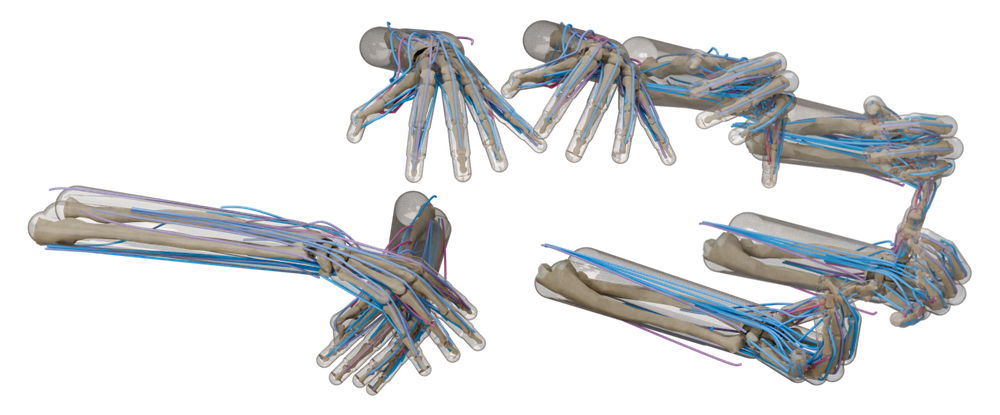}
    \caption{Examples of our tracking policy reproducing long-tail distributed, uncommon poses and trajectories in the reference dataset. The shown examples include challenging behaviors of glissando, where the hand slides across the keyboard, wrist supination, and hand clasping.}
    \label{fig:track_perform}
\end{minipage}\hfill
\begin{minipage}{\columnwidth}
    \centering
    \includegraphics[width=0.32\linewidth]{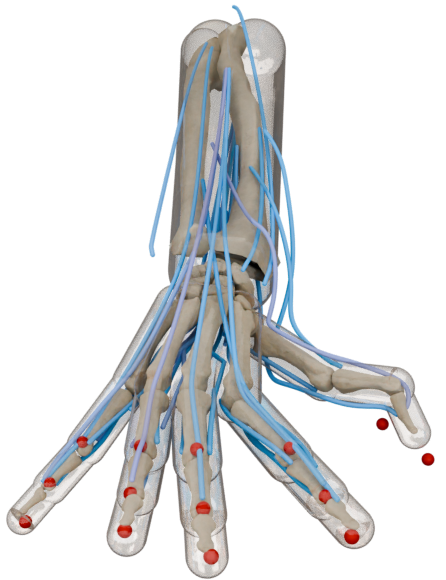}
    \hfill
    \includegraphics[width=0.32\linewidth]{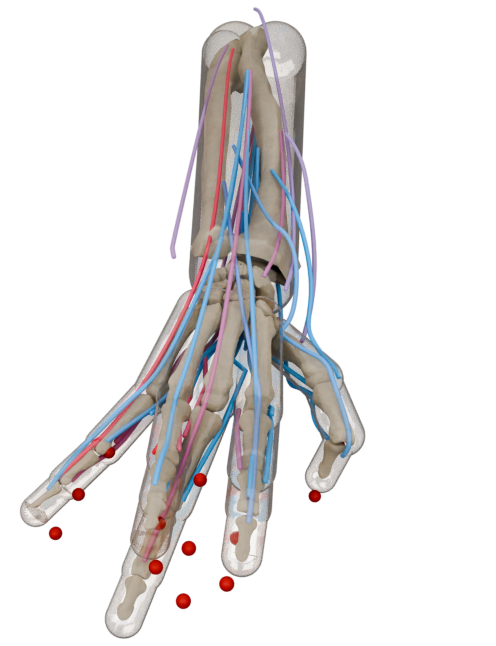}
    \includegraphics[width=0.32\linewidth]{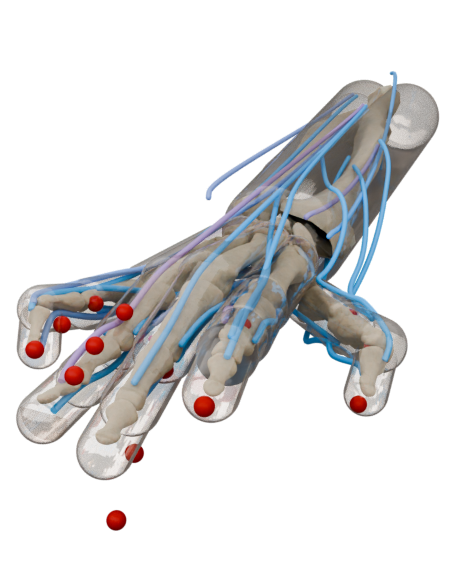}\\
    \includegraphics[width=0.32\linewidth]{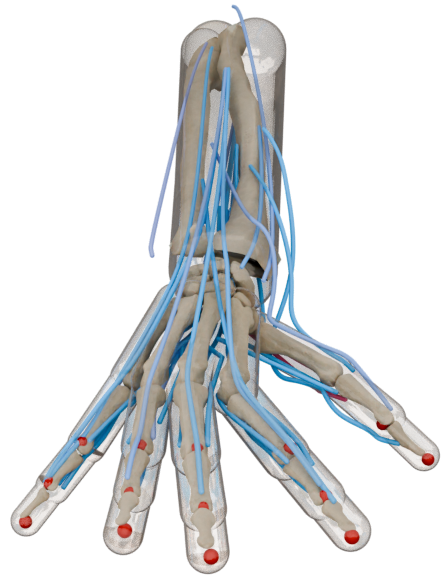}
    \hfill
    \includegraphics[width=0.32\linewidth]{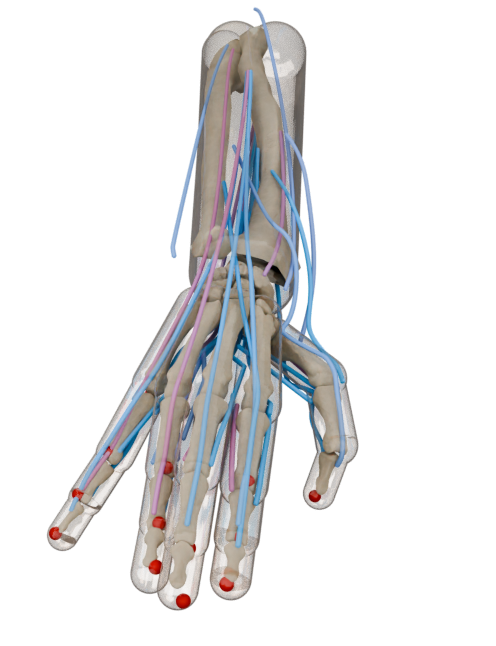}
    \includegraphics[width=0.32\linewidth]{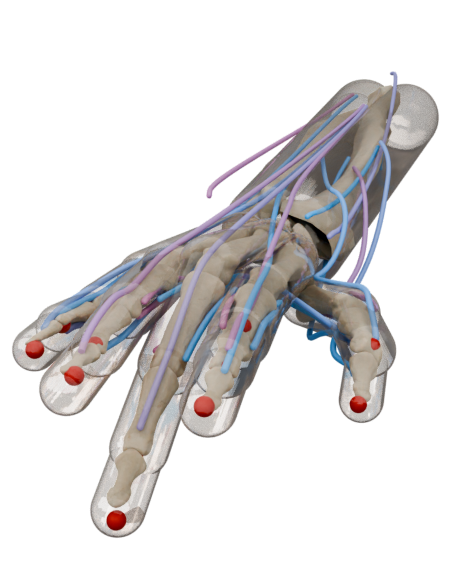}
    \caption{Comparison of Myohand (top) and our enhanced hand model (bottom) while tracking challenging poses with extreme adduction (squeezing), abduction (spreading), and flexion (finger pressing). The red spheres indicate the tracking targets of finger joints and tips.}
    \label{fig:track_comparison}
\end{minipage}
\end{figure*}

We first show our general low-level tracking policy $\pi_{\text{track}}$ capable of tracking all motions in the long-tail distributed reference dataset.
The length of tracking trajectories varies from 20s (1,200~frames) to 18m (64,742~frames).
Such long trajectory lengths impose an additional challenge on motion tracking, since the tracking error would accumulate. 
The performance of the tracking policy $\pi_\text{track}$ using our enhanced hand model is illustrated in Table~\ref{tab:track_err} (Ours). The tracking policy achieves an average error of less than 4mm for all end-effector fingertips and the wrists on this challenging dataset. 
In Fig.~\ref{fig:track_perform}, we visualize tracking results of challenging behaviors in the reference dataset.
By incorporating adaptive sampling, the tracking policy is able to spend more training on those underrepresented and poorly tracked trajectories, enabling accurate tracking across the entire dataset. 
A quantitative ablation study regarding the adaptive sampling is provided in the supplementary materials.

\vspace{\baselineskip}\noindent\textbf{Evaluation on Musculoskeletal Hand Models.} 
To demonstrate the effectiveness of our enhanced hand model, we train an additional tracking policy using identical training setups, but the original MyoHand. The quantitative results are summarized in Table~\ref{tab:track_err}.
As shown in the table, the policy trained with MyoHand exhibits larger tracking errors on the reference dataset, particularly for the thumb and pinky fingertips, where errors exceed 1 cm and 5 mm, respectively.

\begin{table}[t]
    \caption{Tracking errors of the wrists and fingertips, averaged over frames. Results are reported as $\text{mean} \pm \text{std}$ in millimeters (mm). In each cell, the top and bottom values correspond to the left and right hands, respectively.}
    \setlength\tabcolsep{0.09cm}
    \centering
    \small
    \begin{tabular}{ccccccc}
    \toprule
    unit: mm & wrist & thumb & index & middle & ring & pinky\\
    \midrule
        \multirow{2}{*}{MyoHand} 
        & $3.7\pm2.2$ & $11.2\pm8.3$ & $4.9\pm6.3$ & $3.9\pm5.9$ & $3.7\pm4.9$ & $6.4\pm6.0$
 \\
        &$2.9\pm1.6$ & $12.1\pm8.0$ & $4.1\pm4.1$ & $3.7\pm4.1$ & $3.1\pm3.2$ & $5.1\pm4.4$

 \\
        \midrule
        \multirow{2}{*}{Ours} & $1.5\pm1.8$ & $2.4\pm4.3$ & $3.2\pm3.3$ & $3.1\pm3.0$ & $3.3\pm3.0$ & $3.5\pm2.6$

 \\
        &$1.4\pm1.4$ & $2.3\pm3.3$ & $3.7\pm3.8$ & $3.1\pm2.7$ & $3.1\pm3.1$ & $3.8\pm2.1$ \\
    \bottomrule
    \end{tabular}
    \label{tab:track_err}
\end{table}

Figure~\ref{fig:track_comparison} compares the models on challenging poses beyond piano playing. MyoHand (top row) often fails under these extreme configurations. Specifically, the PIP and DIP joints of the thumb and pinky frequently over-bend and fail to return to a neutral configuration after strong abduction (left column) and flexion (right column).
MyoHand also displays locking phenomena 
where a joint remains stuck in near-full flexion (middle column).
In contrast, our enhanced model reaches target postures naturally and accurately without joint locking or pathological configurations, demonstrating superior biomechanical stability and controllability.
We refer to the supplementary video for animated results.

\begin{figure}
    \centering
    \includegraphics[width=\linewidth]{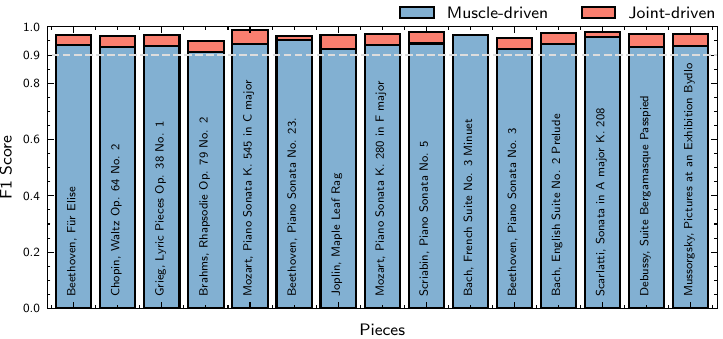}
    \caption{F1 scores of our muscle-driven control policy on the 15 testing pieces~(blue). As a baseline, the performance of control policies trained with a joint-driven hand model actuated through PD servos is shown in red.}
    \label{fig:f1}
\end{figure}

\begin{figure*}
    \centering
    \vspace{-2em}
    \includegraphics[width=0.195\linewidth]{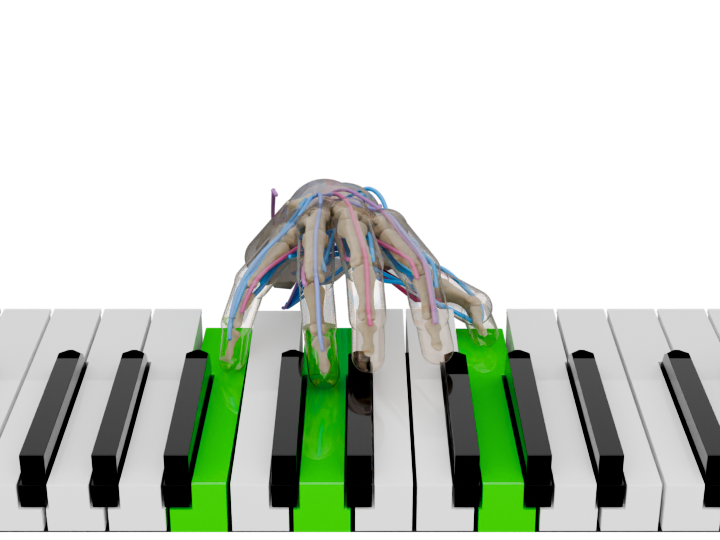}\hfill\includegraphics[width=0.195\linewidth]{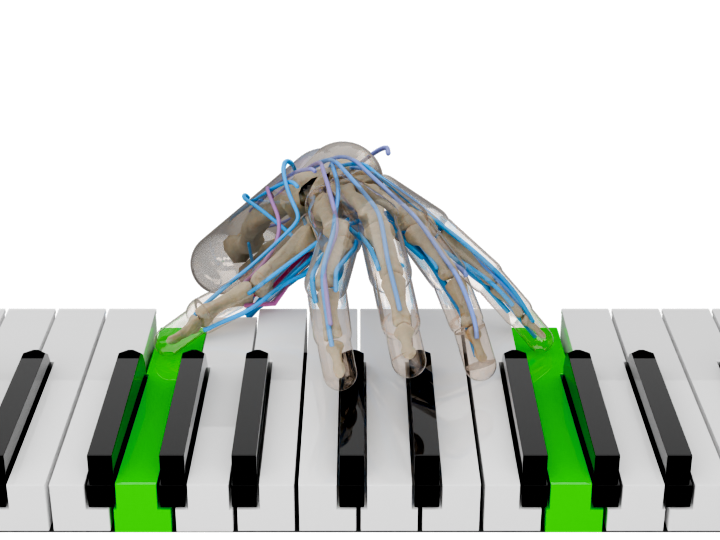}\hfill\includegraphics[width=0.195\linewidth]{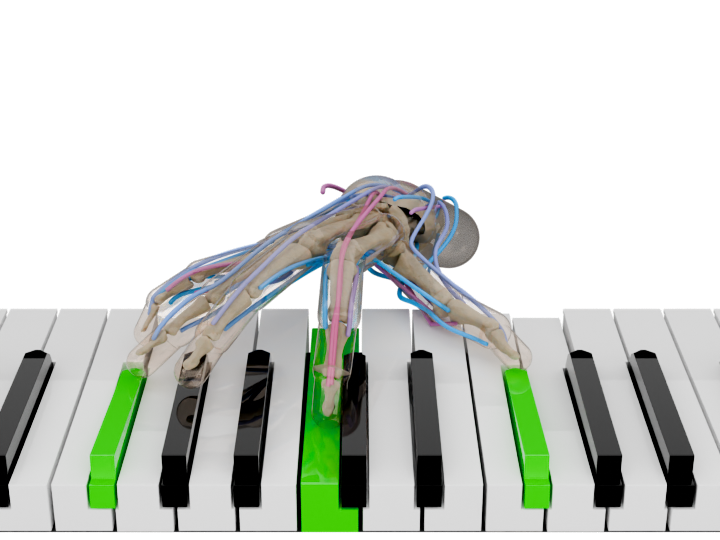}\hfill\includegraphics[width=0.195\linewidth]{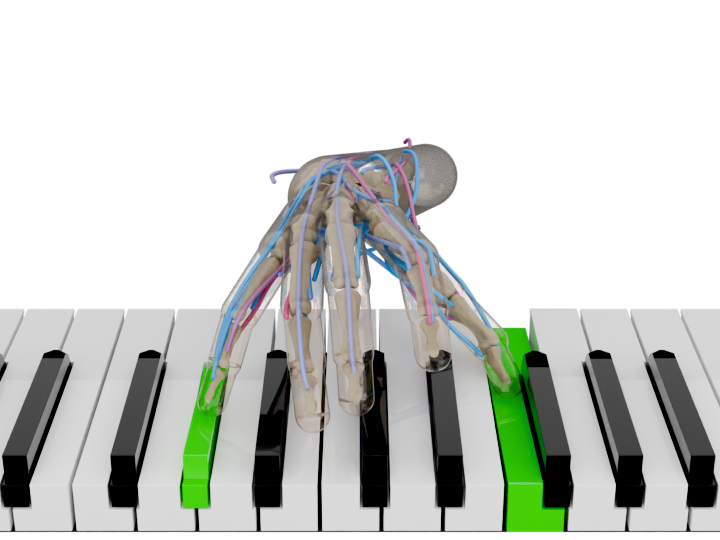}\hfill\includegraphics[width=0.195\linewidth]{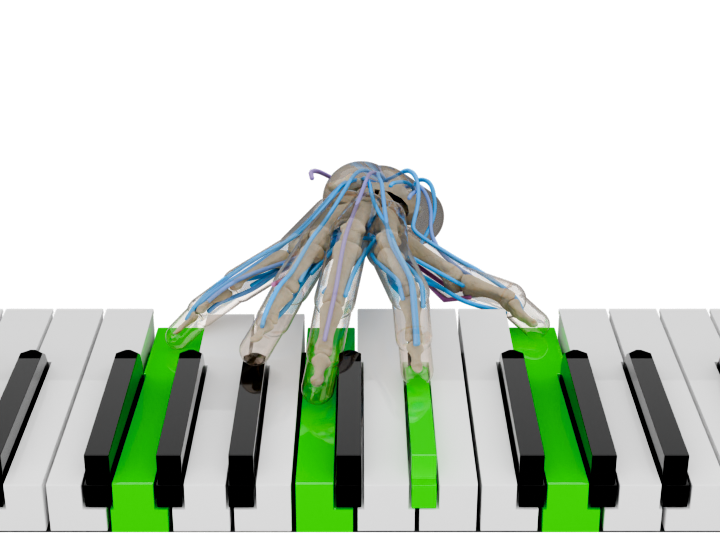}
    
    \vspace{-1em}
    \includegraphics[width=0.195\linewidth]{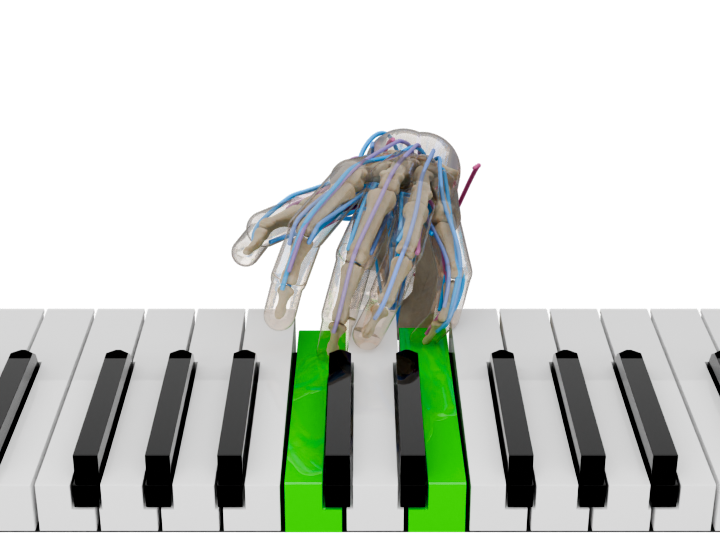}\hfill\includegraphics[width=0.195\linewidth]{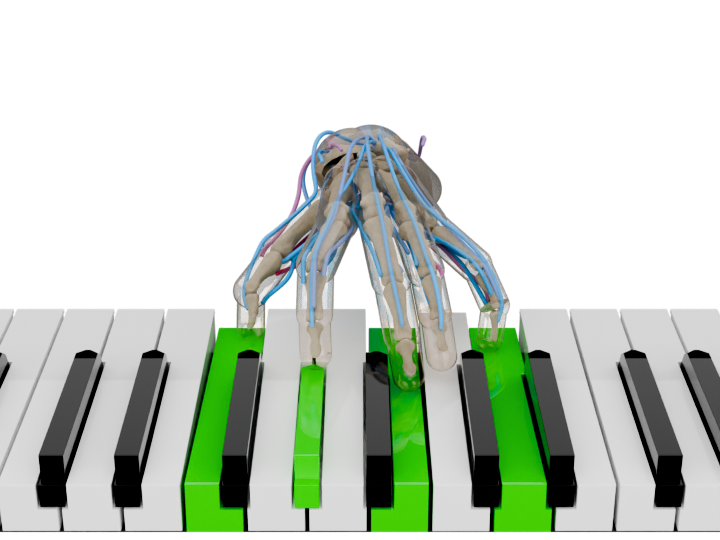}\hfill\includegraphics[width=0.195\linewidth]{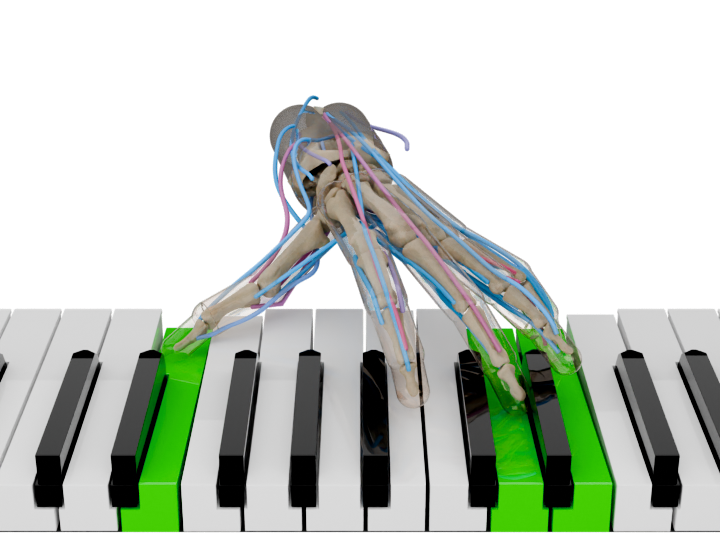}\hfill\includegraphics[width=0.195\linewidth]{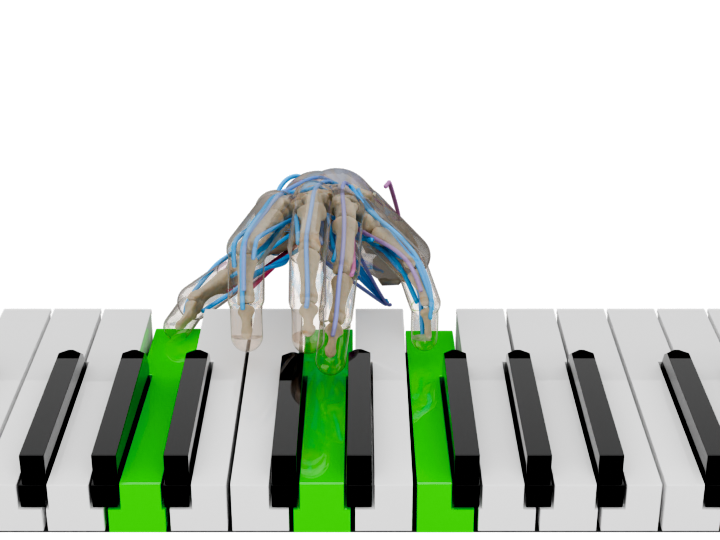}\hfill\includegraphics[width=0.195\linewidth]{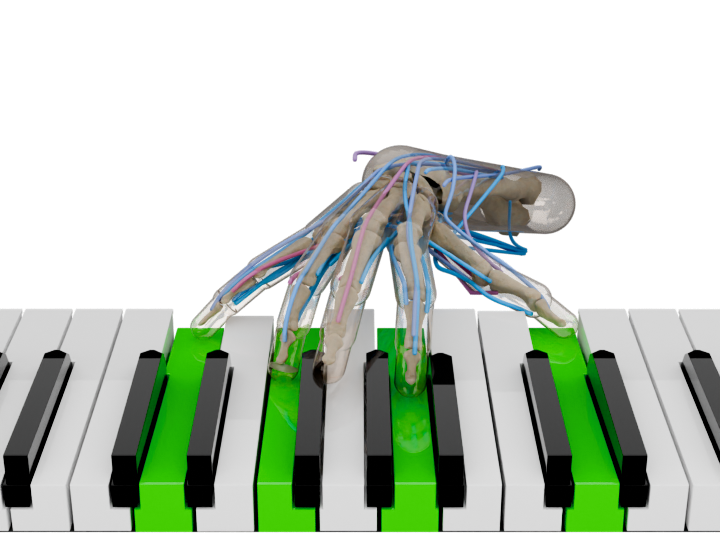}

    \caption{Diverse finger poses when multiple keys are pressed at the same time. Our testing repertoire challenges the system with diverse dexterous behaviors, such as rapid finger alternation (e.g., \textit{Chopin, Waltz Op. 64 No. 2}), sustained chordal control and voicing (e.g., \textit{Brahms, Rhapsodie Op. 79 No. 2}, and \textit{Mussorgsky, Pictures at an Exhibition}), large-scale arm–hand coordination (e.g., \textit{Beethoven, Piano Sonata No.3, 2nd mov.}).
    }
    \label{fig:finger_pose}
\end{figure*}

\subsection{High-Level Motion Synthesis}
We evaluate our approach on 15 music scores excerpted from pieces outside the reference dataset to assess its ability to synthesize bimanual motions for novel pieces.
All evaluated scores are listed in Fig.~\ref{fig:f1}.
Their durations range from 15s to 32s, with note counts varying from 100 to 354.
Together, they span a broad range of musical styles and technical demands, including arpeggios, hand overlapping, and large leaps.

\begin{figure*}
    \centering
    \includegraphics[width=0.159\linewidth]{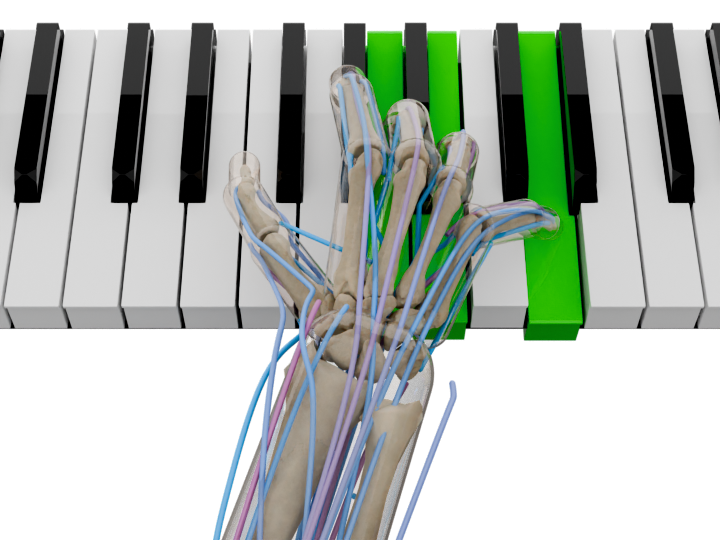} 
    \includegraphics[width=0.159\linewidth]{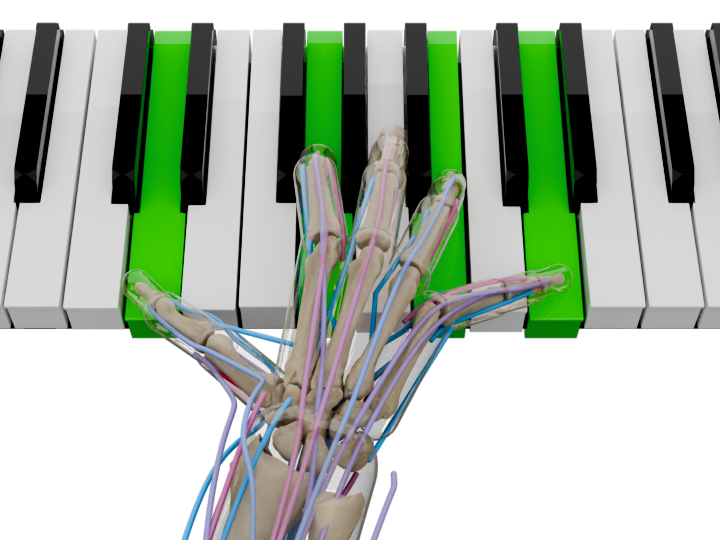}\hfill
    \includegraphics[width=0.159\linewidth]{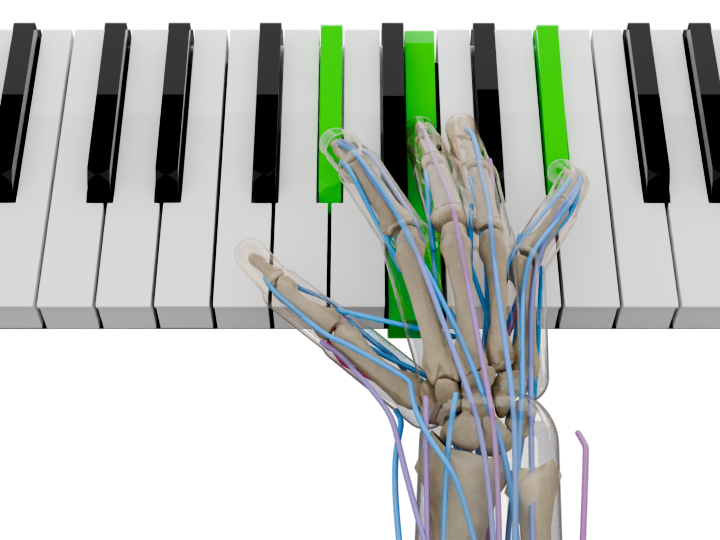} 
    \includegraphics[width=0.159\linewidth]{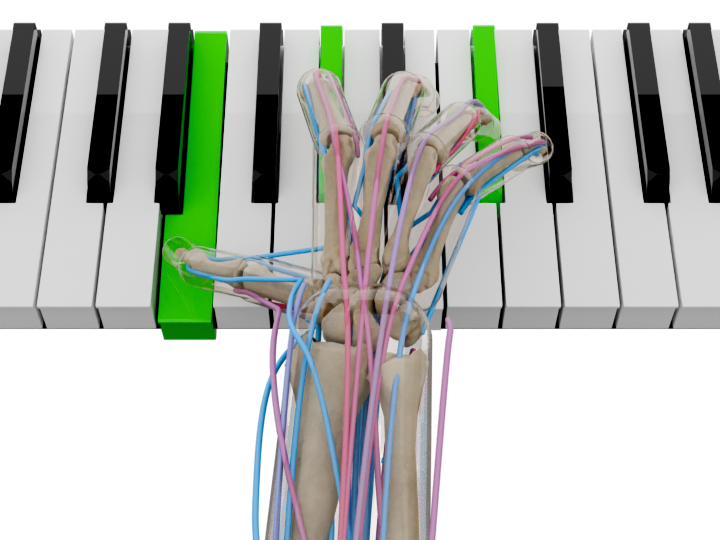}\hfill
    \includegraphics[width=0.159\linewidth]{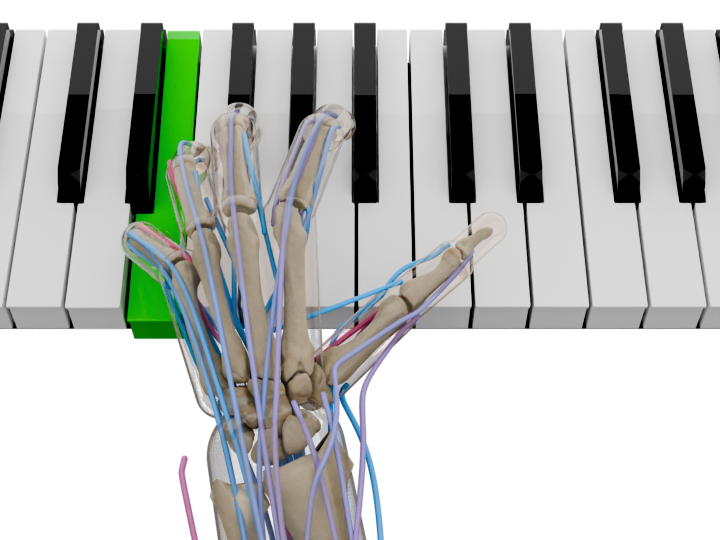} 
    \includegraphics[width=0.159\linewidth]{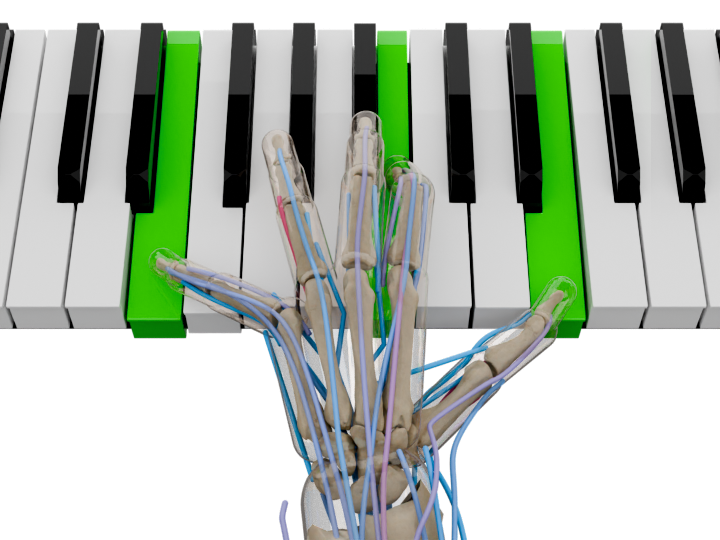}

    \caption{Our policy can infer fingering while taking into account finger occupancy for future key-pressing targets. Two screenshots in each group illustrate the key-pressing states for two consecutive targets.
    Our policy can ensure that all the targets are pressed correctly without the common keys being released.}
    \label{fig:finger_infer}
\end{figure*}

\begin{figure*}
    \centering
    \includegraphics[width=0.195\linewidth]{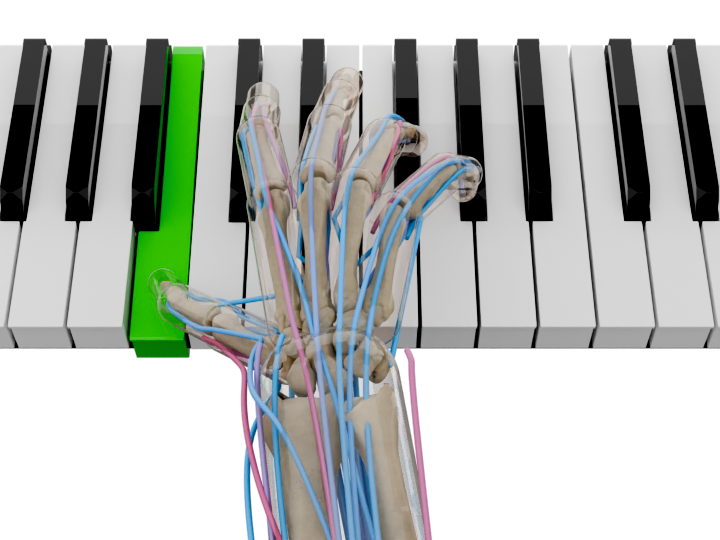}\hfill
    \includegraphics[width=0.195\linewidth]{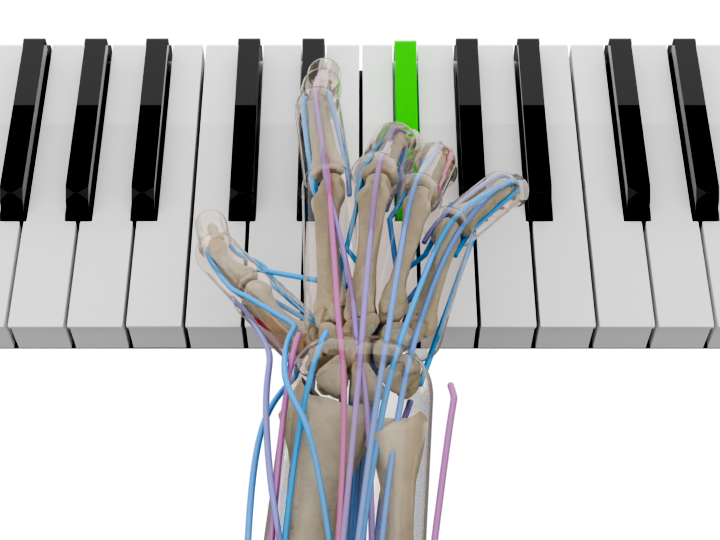}\hfill
    \includegraphics[width=0.195\linewidth]{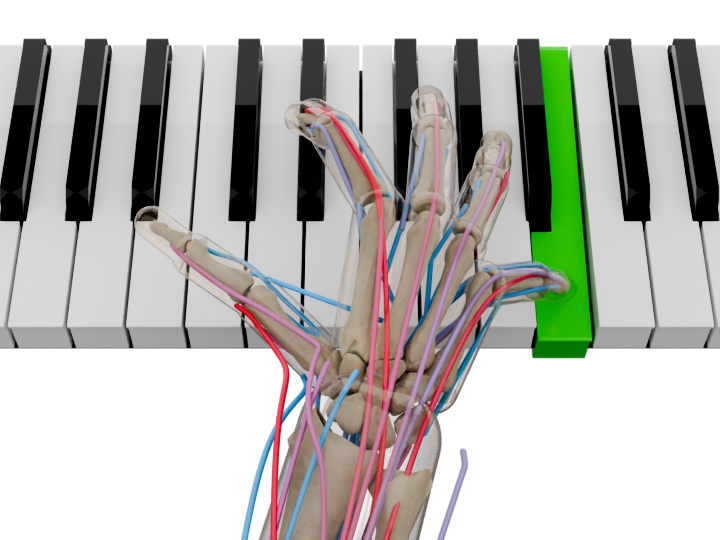}\hfill
    \includegraphics[width=0.195\linewidth]{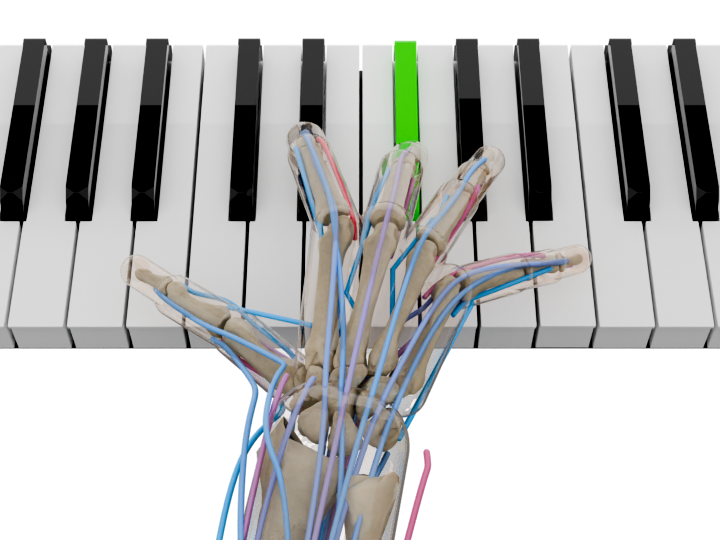}\hfill
    \includegraphics[width=0.195\linewidth]{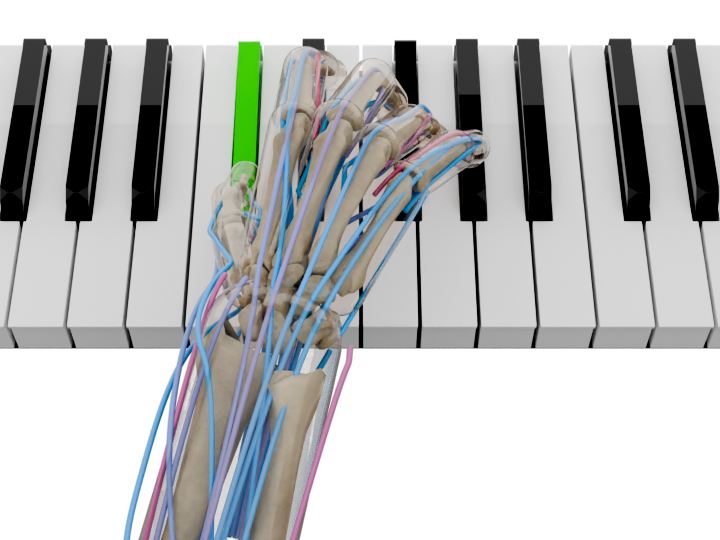}
    
    \includegraphics[width=0.195\linewidth]{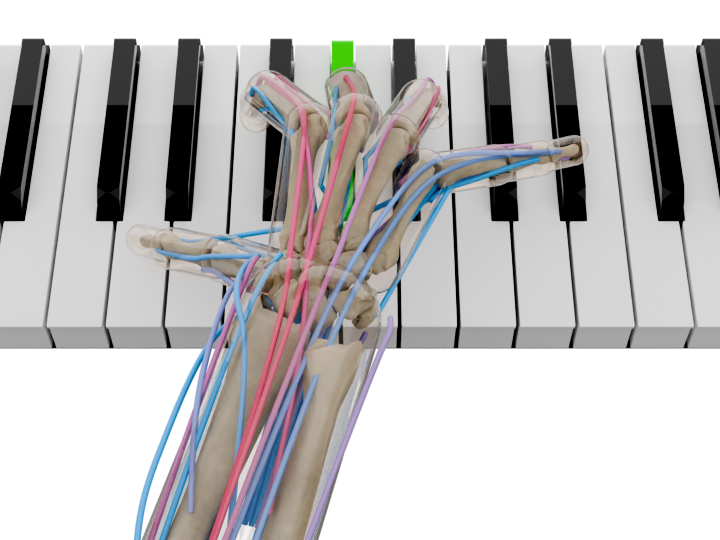}\hfill
    \includegraphics[width=0.195\linewidth]{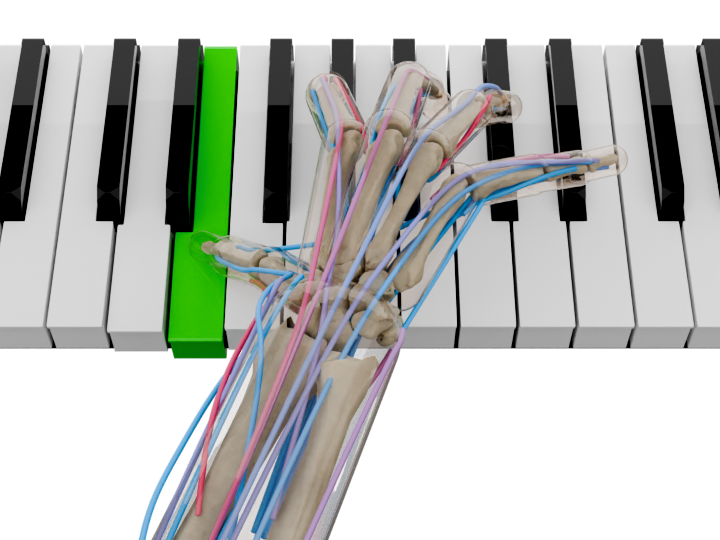}\hfill
    \includegraphics[width=0.195\linewidth]{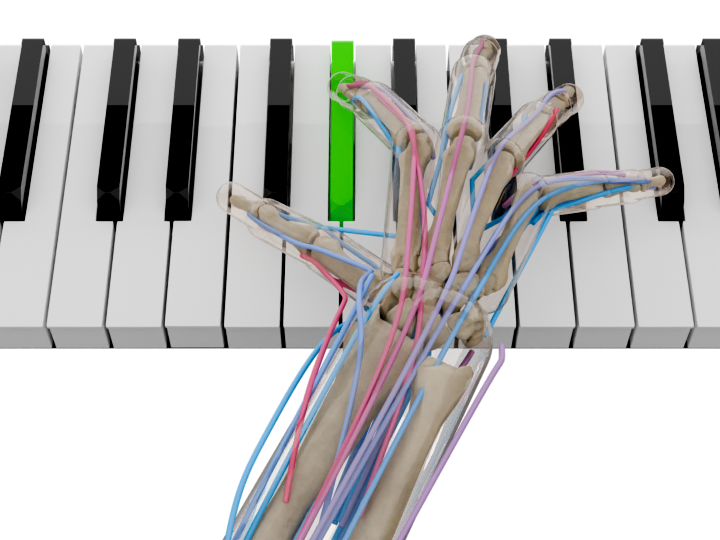}\hfill
    \includegraphics[width=0.195\linewidth]{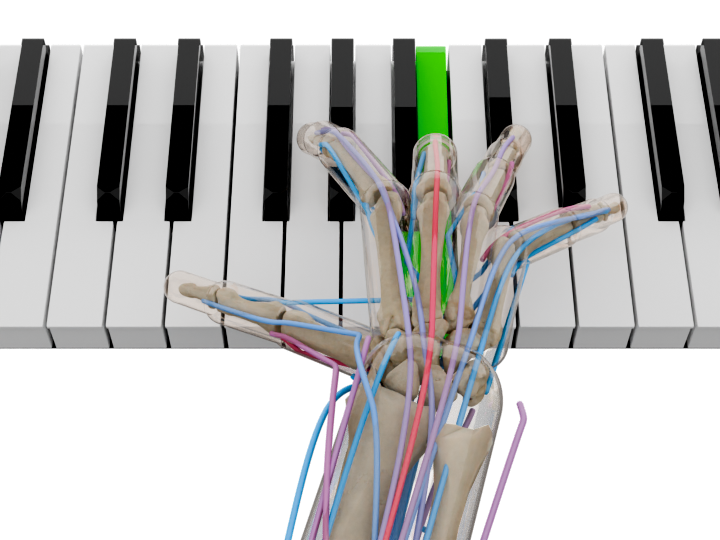}\hfill
    \includegraphics[width=0.195\linewidth]{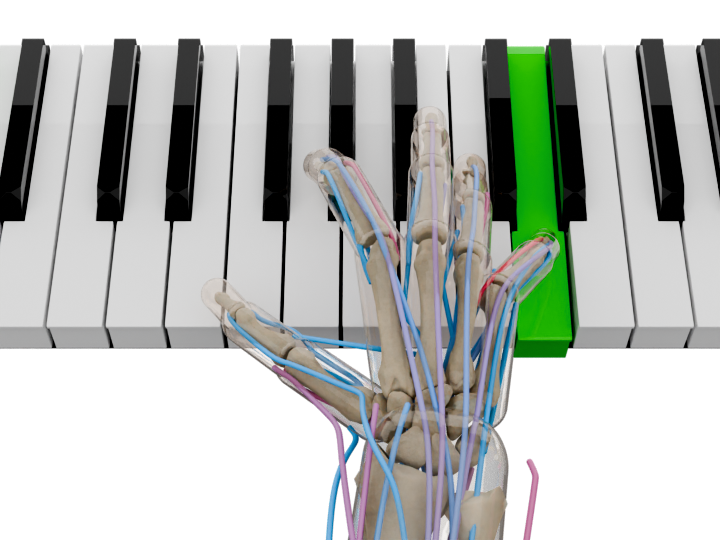}
    \caption{Finger poses during playing arpeggios of \textit{Bach} suites.
    The top row shows an example from \textit{French Suite}, and the bottom row from \textit{English Suite}. %
    }
    \label{fig:arpeggios}
\end{figure*}

\vspace{\baselineskip}\noindent\textbf{Metrics and Baselines.} 
We report F1 score in Fig.~\ref{fig:f1}. To contextualize these results, we compare against a joint-driven baseline that shares the same morphological definition and joint configuration as our musculoskeletal model, while adopting the identical training setups in Sec.~\ref{sec:high}. Given its reduced actuation complexity, this baseline represents the performance ceiling of the high-level controllers.

\vspace{\baselineskip}\noindent\textbf{Results.} As shown in the Fig.~\ref{fig:f1}, the joint-driven baseline achieves near-perfect performance, with F1 scores above 0.96 for all tested scores.
Our muscle-driven model attains F1 scores above 0.9 across all pieces, with an average score of 0.94. 
Despite the increased complexity introduced by muscle dynamics, it matches the performance of the joint-driven model on several pieces, including \textit{Beethoven, Piano Sonata No. 23}, \textit{Bach, French Suite No. 3 Minuet}, and \textit{Scarlatti, Sonata in A major K. 208}. These results demonstrate the effectiveness of our approach for motion synthesis and muscle-driven control.

\begin{figure*}
    \centering
    \includegraphics[width=0.195\linewidth]{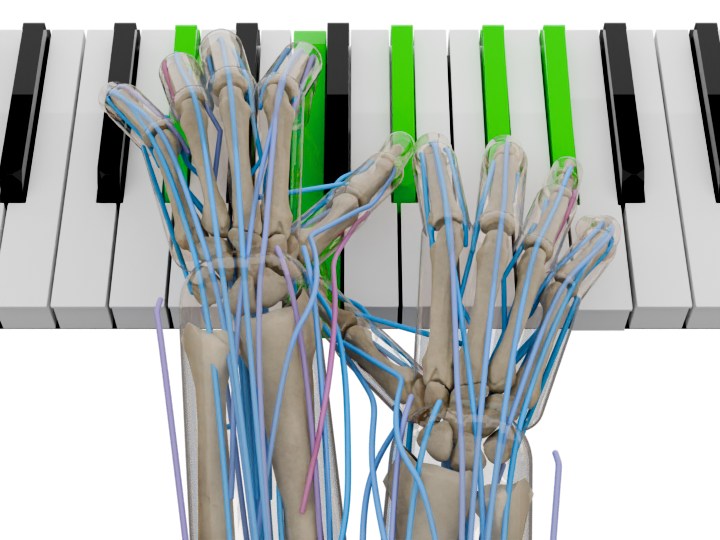}\hfill
    \includegraphics[width=0.195\linewidth]{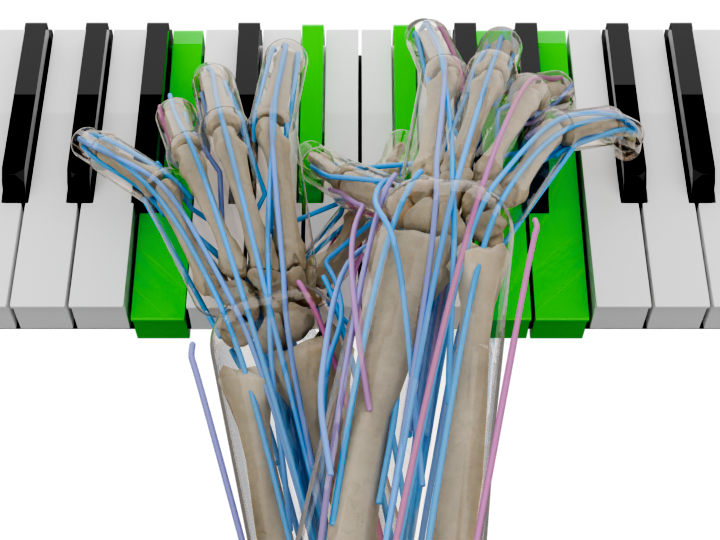}\hfill
    \includegraphics[width=0.195\linewidth]{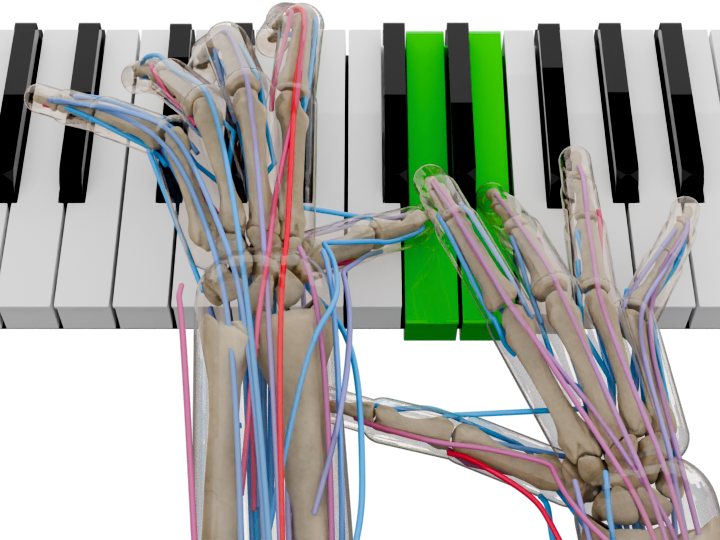}\hfill
    \includegraphics[width=0.195\linewidth]{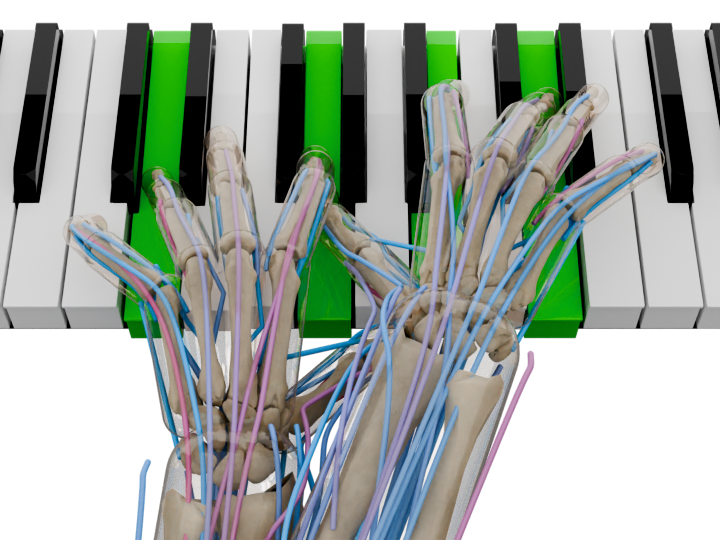}\hfill
    \includegraphics[width=0.195\linewidth]{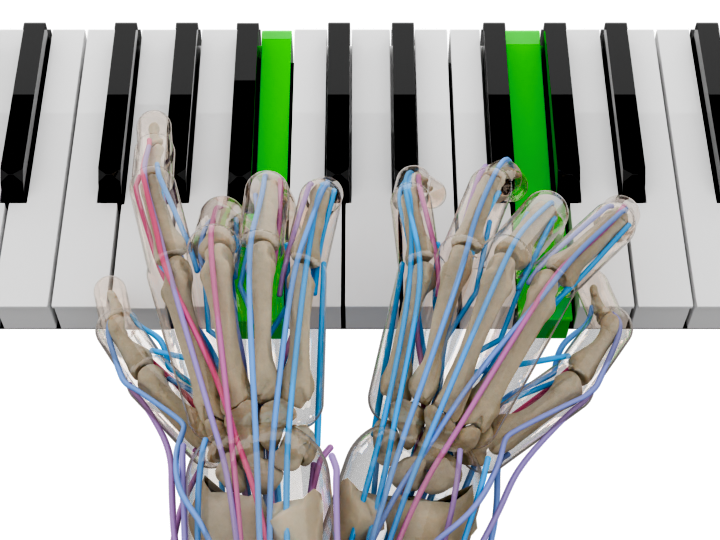}\hfill
    \caption{Two-hand coordination with overlapping and crossover (the rightmost).}
    \label{fig:overlap}
\end{figure*}

\begin{figure*}
    \centering
    \includegraphics[width=.325\linewidth]{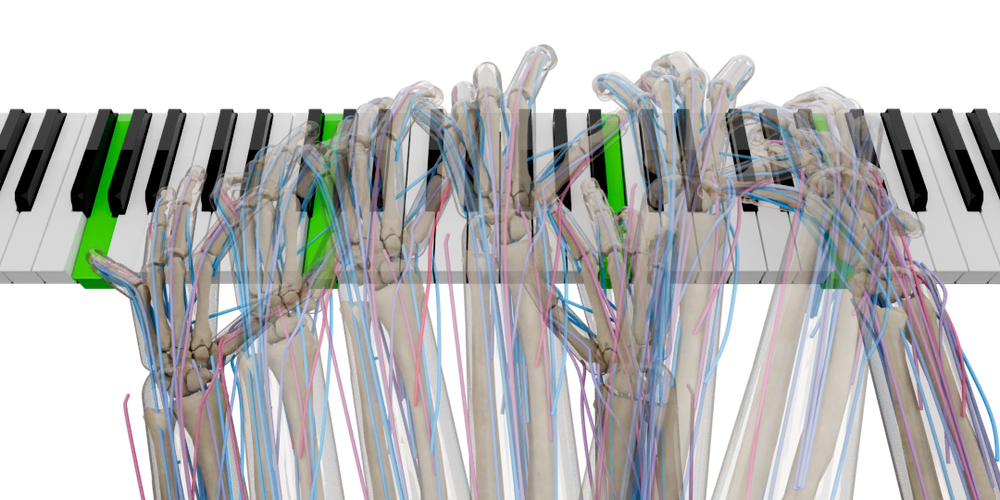}\hfill
    \includegraphics[width=.325\linewidth]{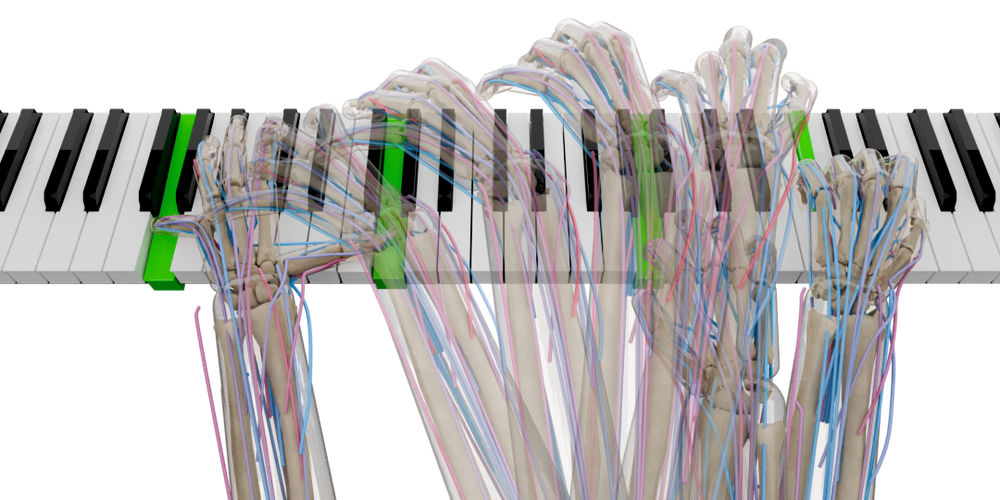}\hfill
    \includegraphics[width=.325\linewidth]{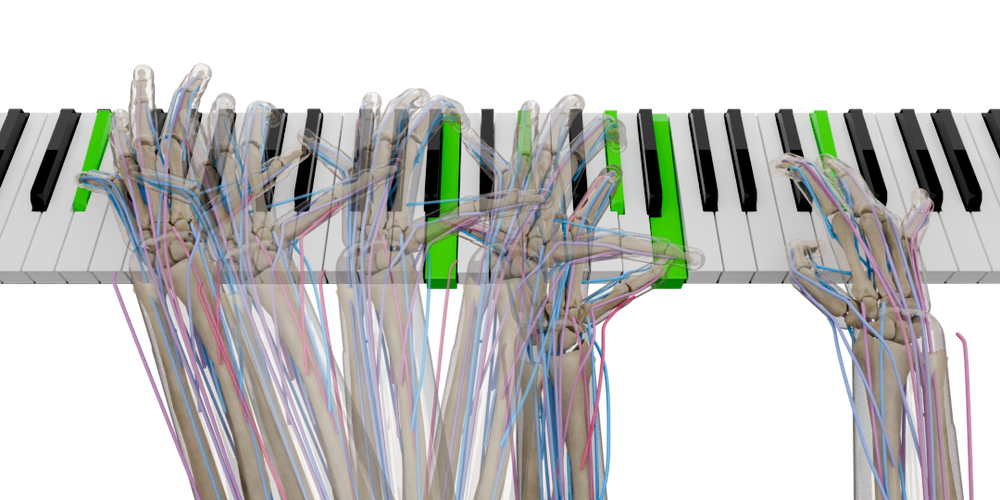}
    \caption{Demonstrations of large hand leaps. The left and middle screenshots show the left hand leaping over the right hand and back.}
    \label{fig:leap}
\end{figure*}

\vspace{\baselineskip}\noindent\textbf{Emergent Behaviors.}
Figures~\ref{fig:teaser} and~\ref{fig:finger_pose} showcase the diverse finger configurations emerging during complex multi-key presses. Our pipeline generates independent finger motions while maintaining stable wrist postures, despite the contact-rich interactions. Notably, as the low-level tracker and latent manifold were trained in open space without a piano, the high-level controller has to learn to manage these contacts entirely through exploration within the latent space.
The policy also demonstrates an emergent ability to infer fingering without explicit supervision (Fig.~\ref{fig:finger_infer}).
Similarly, during arpeggios (Fig.~\ref{fig:arpeggios}), the policy naturally selects optimal fingering to minimize wrist movement while traversing sequential notes.
Though trained through a decentralized multi-agent setting,
the high-level controller successfully synthesizes challenging coordinated bimanual behaviors. In scenarios where hands spatially overlap (Fig.~\ref{fig:overlap}), the policies resolve potential inter-hand collisions to maintain %
effective pressing. Even during rapid, large-scale leaps across the keyboard (Fig.~\ref{fig:leap}), the model demonstrates the spatial awareness and timing precision necessary to accurately localize and strike target keys.

\subsection{Model Validation}\label{sec:exp_hand_model}

We evaluate the physiological fidelity of our model by comparing muscle activation patterns generated by the tracking policy with electromyography (EMG) recordings from a human subject.

\begin{figure}
    \centering
    \includegraphics[width=0.49\linewidth]{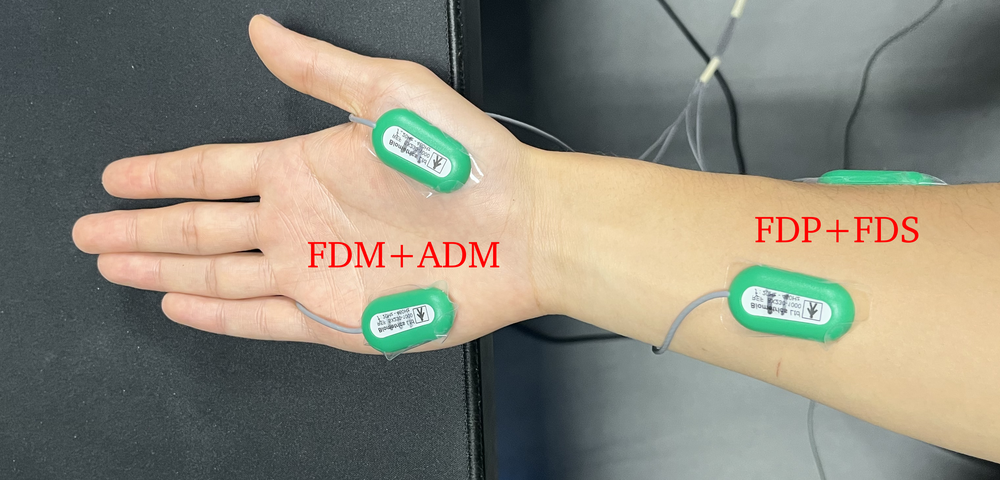}
    \hfill\includegraphics[width=.49\linewidth]{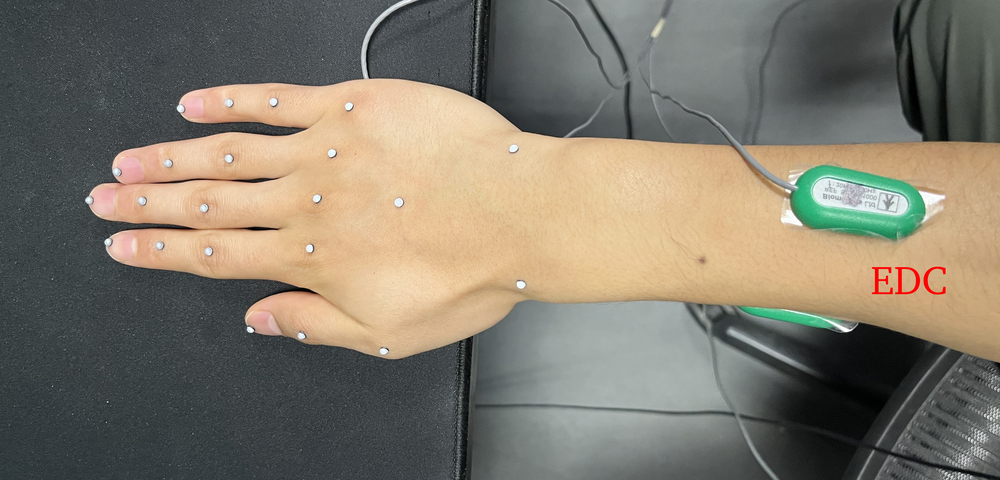} \\
    
    \vspace{0.5cm}
    \includegraphics[width=\linewidth]{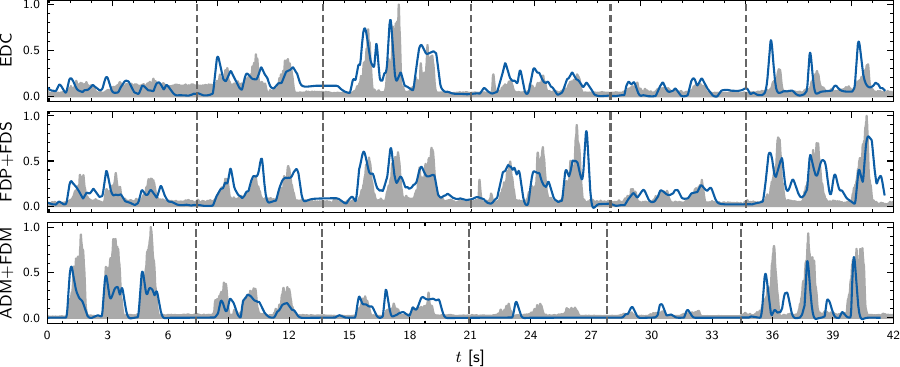}
    \caption{Comparison of EMG recordings collected from a human subject (shadow regions) and the activations (blue lines) generated by our model during motion tracking. Top: illustrations of the EMG sensor placement.}
    \label{fig:emg}
\end{figure}

\vspace{\baselineskip}\noindent\textbf{Experimental Setup.} 
In the top row of Fig.~\ref{fig:emg}, we show the EMG sensor placement.
For intrinsic muscles (which originate and insert within the hand), we target the ADM and FDM group, as these are superficial and accessible. For extrinsic muscles (located in the forearm), we measure the EDC and FDP+FDS groups. 
Given the anatomical proximity of these muscles, we assume surface EMG measurements reflect their combined activity rather than distinct, muscle-specific signals. During data collection, the subject performs a sequence of tasks, including isolated finger flexion/extension, hand abduction, and key pinches, repeating each task three times.
We record hand kinematics and EMG signals simultaneously to ensure temporal alignment between motion and muscle activity.

\vspace{\baselineskip}\noindent\textbf{Results.} To ensure the validity of the comparison, we first verify that the tracking policy faithfully reproduces the subject's motion. The policy achieves high accuracy, with average tracking errors of 0.1 for the wrist and 0.9, 2.5, 2.6, and 1.8 for the thumb, index, middle, ring, and pinky fingertips, respectively, in the unit of mm.
We normalize the raw, noisy EMG signals using standard post-processing, including full-wave rectification and low-pass filtering.
Fig.~\ref{fig:emg} displays the processed EMG measurements (gray shade) alongside the activation generated by the tracking policy (blue curve). The model-generated activations are consistent with the overall temporal patterns of the measured EMG signals. This consistency aligns with the underlying motion tasks (separated by dashed lines): from left to right, the phases correspond to flexion/extension of the pinky, ring, middle, and index fingers, followed by an idle period, and finally hand spreading and squeezing. The three consecutive repetitions of each motion appear as three distinct peaks in the generated activations.
While we observe minor discrepancies in certain segments—expected due to inter-subject anatomical variability, habitual motor strategies, and the inherent limitations of surface EMG—the strong agreement in activation trends supports the physiological relevance of our proposed muscle-driven controller.

\section{Sensitivity Analysis}
In this section, we conduct a series of sensitivity analyses to evaluate the key components within our approach.

\subsection{Adaptive Sampling}
Table~\ref{tab:track_err_noadpt} compares the tracking performance of policies trained on our enhanced musculoskeletal hand model with and without the proposed adaptive sampling strategy. 
All policies are trained using identical network architectures and hyperparameters, differing only in whether adaptive sampling is employed. 
The results show that incorporating the MCMC-based adaptive sampling substantially improves tracking accuracy across the entire reference dataset. 
Without adaptive sampling, the policy exhibits average tracking errors on the order of centimeters with large standard deviations, indicating poor coverage of challenging motion segments. 
In contrast, adaptive sampling effectively focuses training on poorly tracked trajectories, reducing the average tracking error to below 4mm and enabling accurate and stable tracking over the full dataset.

\begin{table}[t]
    \setlength\tabcolsep{0.05cm}
    \centering
    \small
    \caption{Performance of the tracking policies trained with adaptive sampling (ours) and those without the adaptive sampling (w/o A.S.)
    The reported numbers are the tracking errors of the wrists and fingertips in the form of mean$\pm$std, averaged over all frames in the reference dataset. In each cell, the top and bottom values correspond to the left and right hands, respectively.}
    \begin{tabular}{ccccccc}
    \toprule
    unit: mm & wrist & thumb & index & middle & ring & pinky\\
    \midrule
        \multirow{2}{*}{Ours} & $1.5\pm1.8$ & $2.4\pm4.3$ & $3.2\pm3.3$ & $3.1\pm3.0$ & $3.3\pm3.0$ & $3.5\pm2.6$

 \\
        &$1.4\pm1.4$ & $2.3\pm3.3$ & $3.7\pm3.8$ & $3.1\pm2.7$ & $3.1\pm3.2$ & $3.8\pm2.1$ \\
        \hline
        \multirow{2}{*}{w/o A.S.} & $16.7\pm4.7$ & $98.3\pm18.3$ & $68.2\pm10.7$ & $22.2\pm8.0$ & $30.8\pm9.5$ & $65.9\pm12.0$ \\
        & $17.0\pm3.4$ & $94.4\pm18.0$ & $74.0\pm12.1$ & $22.8\pm6.4$ & $19.3\pm6.0$ & $41.6\pm13.8$ \\
    \bottomrule
    \end{tabular}
    \label{tab:track_err_noadpt}
\end{table}

\begin{figure}
    \centering
     \includegraphics[width=\linewidth]{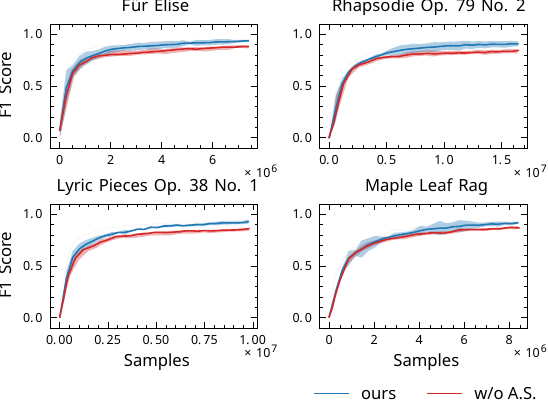}
    \caption{Learning performance of the high-level controllers with adaptive sampling (ours) and that using uniform sampling (w/o A.S.).}
    \label{fig:ablation_ada}
\end{figure}

Figure~\ref{fig:ablation_ada} illustrates the learning performance, measured by F1 scores, when adaptive sampling is used during high-level training, compared to uniform sampling. Adaptive sampling biases training toward poorly performed notes, which may lead to lower running performance during training but results in improved evaluation performance at test time, as shown in the figure. Overall, during high-level controller training, adaptive sampling improves convergence speed by approximately 5–10\%, particularly in the later stages of training when only a small subset of notes remains challenging.

\subsection{Control Space}

\begin{figure}
    \centering
    \includegraphics[width=\linewidth]{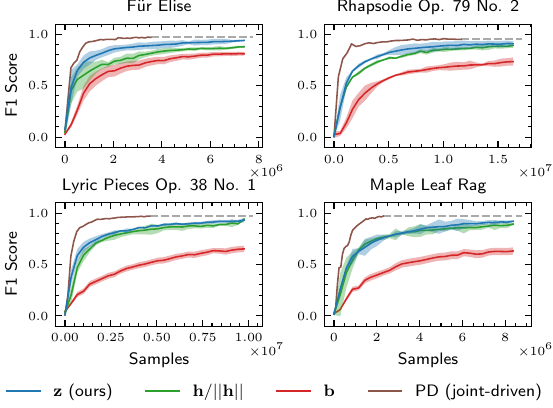}
    \caption{Learning performance of the high-level controllers while using different control spaces. $\textsc{PD}$ refers to the baseline using a joint-driven hand model. The joint-driven model has the same morphology and joint configurations as our musculoskeletal model, but is actuated by PD servos.}
    \label{fig:ablation_space}
\end{figure}

\begin{figure}
    \hfill
    \includegraphics[width=\linewidth]{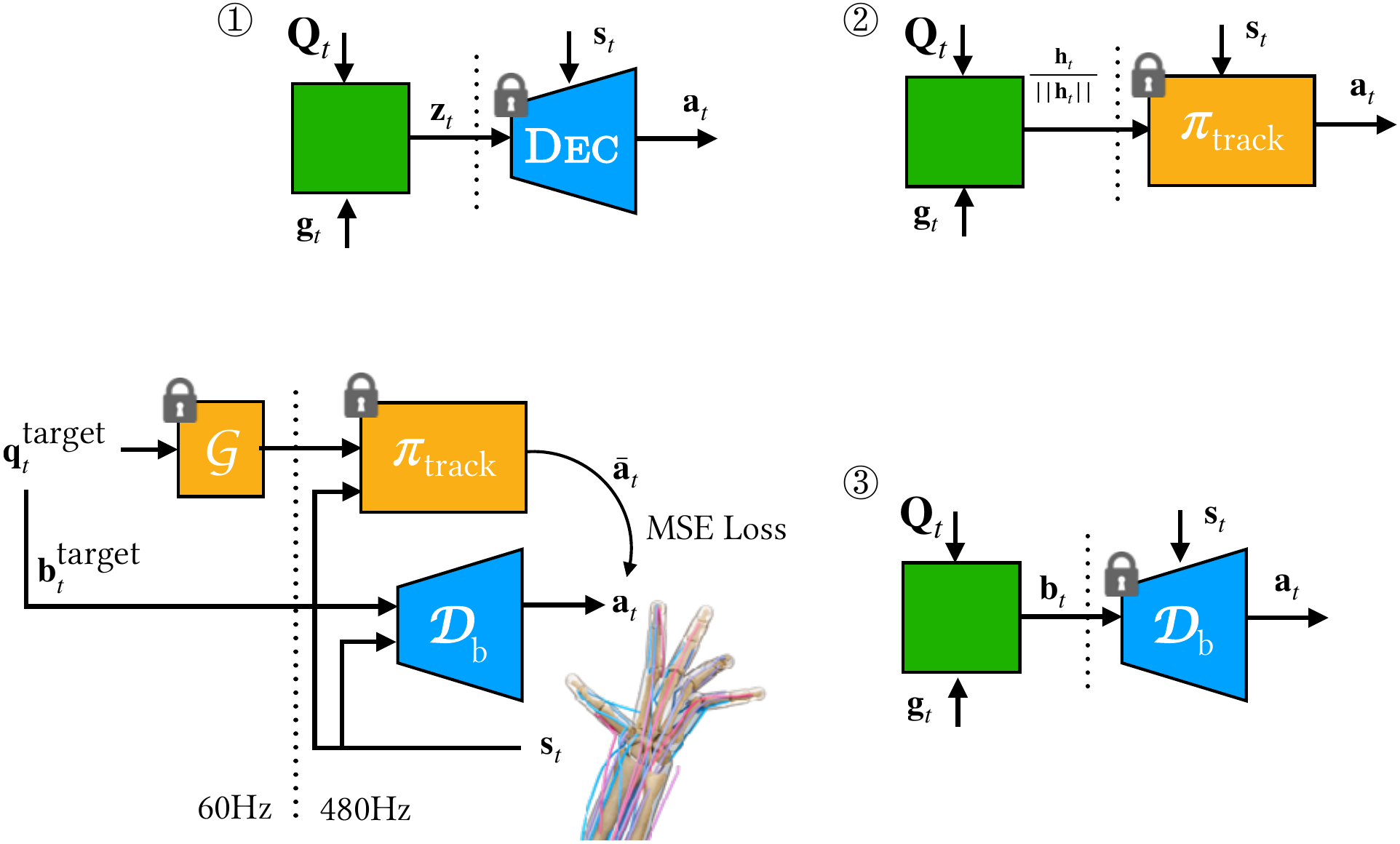}
    \caption{Network architectures with different latent spaces. From \textcircled{1} to \textcircled{3}, the latent spaces are (1) the VAE latent used in our approach, (2) the goal embedding of the tracking policy, constrained on the surface of a unit sphere, and (3) an explicit space defined by the joint pose $\mathbf{b}_t$, respectively.
    Additionally, the subfigure at the bottom left illustrates the diagram to distill a tracking policy into an explicit joint-pose space. While $q_t^\text{target}$ includes the target poses of body links in $H$ frames, $\mathbf{b}_t^\text{target}$ only contains the immediate target joint pose at timestep $t$. We follow the dual-frequency hierarchical architecture, and let the high-level controllers for motion synthesis (green boxes) run at a lower frequency of 60Hz to ensure the training efficiency.
    }
    \label{fig:net_cspace}
\end{figure}

Figure~\ref{fig:ablation_space} compares learning performance in terms of F1 scores when different control spaces are used to train the high-level controller. 
The corresponding network architectures are shown in Fig.~\ref{fig:net_cspace}. 
We omit the baseline that directly uses the tracking latent $\mathbf{h}$, as this representation does not support effective high-level policy learning in our experiments. 
Instead, we introduce a variant of the tracking policy that uses a normalized latent $\mathbf{h}/||\mathbf{h}||$ as the goal representation. 
This normalization constrains the latent to the surface of a unit hypersphere $\mathbb{S}^{\dim \mathbf{h}}$, yielding a space with more densely and uniformly distributed valid points.
Owing to our dual-frequency architecture, the high-level controller can operate over $\mathbf{h}/\lVert \mathbf{h} \rVert$ at a low frequency while directly leveraging the tracking policy as a low-level muscle controller without an explicit distillation stage.
We additionally evaluate a baseline in which the control space is defined explicitly at the joint level. 
The objective of this baseline is to approximate a PD-servo-like controller for muscle-driven systems, while taking an immediate target joint pose $\mathbf{b}_t^{\text{target}}$ for the next frame as input.
For reference, we also include a joint-driven hand model actuated by PD servos in the comparison.

As illustrated in the figure, both our proposed approach using the VAE latent $\mathbf{z}$ and the normalized latent $\mathbf{h}/||\mathbf{h}||$ achieve strong performance in high-level policy training, with the VAE latent yielding slightly higher F1 scores overall. 
While quantitative performance is comparable, we observe that motions generated through $\mathbf{h}/||\mathbf{h}||$ often exhibit noticeable jitter, whereas those synthesized through the VAE latent are substantially smoother and more stable. We refer to the supplementary video for visual comparisons.

The joint-space baseline using $\mathbf{b}$, however, performs poorly. 
Although we apply the same on-policy distillation strategy, the resulting low-level controller $\mathcal{D}_b$ exhibits limited generalization.
While it can track reference trajectories with moderate accuracy, it frequently loses control of the hand when driven by stochastic actions provided by the high-level controller.
In addition, analogous to PD servos, we also experimented with a tracking policy that conditions only on a one-frame future target pose.
However, this setting results in jittery motions and large tracking errors.

We note that although the number of training samples required for high-level policy learning is not quite large, muscle–tendon simulation is substantially more expensive than joint-driven simulation.
In our experiments, muscle-driven simulation is approximately twice as slow as joint-driven PD control, leading to a high-level policy training of roughly one week on an NVIDIA RTX Pro 6000 GPU.

\begin{figure}
    \centering
     \includegraphics[width=\linewidth]{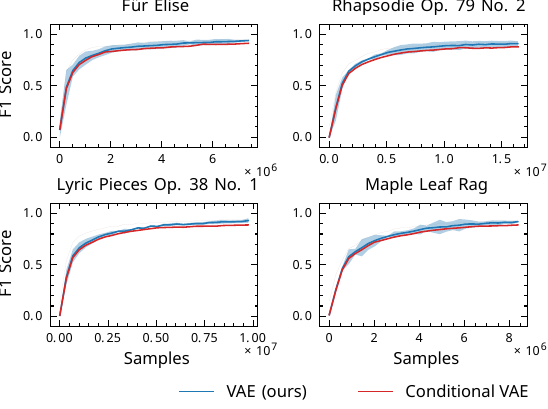}
    \caption{Learning performance of the high-level controllers using a vanilla VAE to perform distillation and that using a conditional VAE.}
    \label{fig:ablation_vae}
\end{figure}
Additionally, Fig.~\ref{fig:ablation_vae} compares the performance of our approach using a vanilla VAE for latent distillation with that of a conditional VAE. In the latter case, we adopt the same distillation objective~(Eq.~\ref{eq:vae}) but replace the prior with a state-conditioned prior $p(\cdot|\mathbf{s}_t)$. 
While prior work~\cite{luouniversal,ling2020character,tessler2024maskedmimic} reports improved performance with conditional VAEs in physics-based control tasks, we observe slightly better results with a vanilla VAE in our setting, along with faster inference due to its simpler architecture.
We attribute this difference to the desynchronized time scales in our hierarchical design: the encoder produces latent codes at a low frequency based on $\mathbf{s}_t$, whereas the decoder operates at a higher frequency to generate muscle activations from instantaneous states based on $\mathbf{s}_{t+\delta}$ (cf. Eq.~\ref{eq:vae}). Introducing a state-conditioned prior in this setting may create a temporal mismatch between latent inference and control execution, which can negatively impact training stability.

\subsection{Decentralized Control vs. Centralized Control}
\begin{figure}
    \centering
     \includegraphics[width=\linewidth]{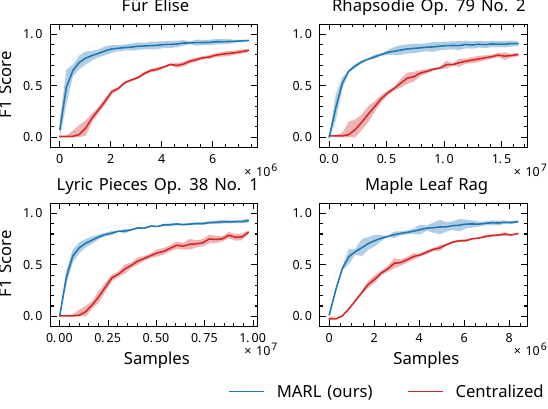}
    \caption{Learning performance of the high-level controllers using a multi-agent setup (MARL) and that using a centralized policy for bimanual control (Centralized). The Centralized setup typically needs around 30\% more training iterations to reach the same final performance as MARL.}
    \label{fig:ablation_marl}
\end{figure}

Figure~\ref{fig:ablation_marl} compares the performance of our decentralized multi-agent reinforcement learning (MARL) formulation with a centralized alternative for high-level policy training.
In the centralized baseline, a single policy network controls both hands jointly, while receiving the same observations as the decentralized policies. 
To ensure a fair comparison, we adopt the same multi-objective learning framework for policy training in the centralized setup, but involving the imitation and task-directed objectives for the two hands at the same time.
As shown in the figure, the decentralized multi-agent formulation yields a clear and consistent advantage over centralized control.
In particular, MARL, with reduced action space complexity, leads to substantially faster convergence during training and achieves higher final performance. 

As described in Sec.~\ref{sec:high}, 
the decentralized setup requires knowing the hand-level key assignment to perform reward computation, though the detailed fingering is unnecessary. 
In the case where the hand-level annotation is unavailable, we can still perform training, using the nearest finger heuristic.
However, the separation of single hands can destabilize training. For example, one hand may avoid all key assignments to maximize its own reward by staying far away from all the target keys.
In such cases, a centralized controller governing both hands yields better performance.

\section{Conclusion}
We present a hierarchical, muscle-driven control system for dexterous piano performance that enables physically simulated hands to synthesize motions for pieces of music outside the reference dataset.
By combining high-frequency muscle-level control with low-frequency latent-space coordination, our approach bridges biomechanically grounded actuation and high-level goal-directed motion synthesis. 
We obtain the operating latent space by distilling a muscle-driven tracking policy into a VAE model.
Over this latent space, the control system can generate diverse and dynamic motions while remaining agnostic to the underlying complexity of muscle–tendon dynamics, resulting in efficient policy training and inference process.
Though focusing on piano playing tasks in this paper, the framework of our approach has the potential to be applied to other forms of muscle-driven dexterous and whole-body control.
We enhance the musculoskeletal hand modeling to provide better support for fine control of fingers.
As demonstrated by our validation against human EMG recordings, our hand model and control policy generate physiologically plausible activation patterns with high fidelity to human motor strategies.
This suggests the possibility of applying our approach to biomechanical analysis of human performance.

As an approach to physics-based dexterous control for the task of piano playing, our framework models the bimanual control problem through multi-agent reinforcement learning, and obtains state-of-the-art performance.
Nevertheless, our approach is still limited to generating binary-represented sound, which is triggered only when a key is fully pressed.
An interesting direction for further work is to improve piano modeling, allowing the generation of realistic sounds that are crucial for expressive performance. For example, we could introduce a physics-based sound generation mechanism or augment the simulation with a key-velocity-based sound model.
We also observe occasional abrupt or stiff movements in the synthesized motions. This behavior may stem from the absence of higher-level constraints, such as muscle fatigue modeling or perception-based velocity control, which could otherwise regularize motion smoothness and temporal consistency.
Although our approach can synthesize motions without requiring detailed fingering annotations, the clef-based heuristic may deviate from optimal hand usage in practice, and we also observe rare cases where the control policy becomes trapped in local minima during fingering inference. Incorporating more advanced heuristics beyond the simple nearest-finger rule could help mitigate these issues.
In addition, the reference dataset does not include forearm motion. During retargeting, we therefore infer elbow poses heuristically (see the supplementary materials for details), which can lead to unnatural elbow configurations that are subsequently imitated by the control policies. Incorporating arm motion capture into the dataset is a promising direction for improving motion naturalness in future work.

\begin{acks}
This work was supported by the Wu-Tsai Human Performance Alliance, the Stanford Institute for Human-Centered Artificial Intelligence (HAI) Google Cloud Credits program, and the 
\grantsponsor{GS100000002}{National Institutes of Health}{http://dx.doi.org/10.13039/100000002}
under Grant No.:~\grantnum{GS100000002}{R01 5R01LM014154-04}. The authors thank Prof. Scott L. Delp, Dr. Jennifer Hicks, and Dr. Nick Bianco (Stanford University) for insightful discussions on hand modeling; Jichao Zhao (Clemson University) for assistance with EMG data collection; and Ruocheng Wang and Haocheng Shi for reproducing the results of Für Elise.
\end{acks}

\bibliographystyle{ACM-Reference-Format}
\bibliography{reference}

\appendix
\renewcommand{\thefigure}{S\arabic{figure}}
\renewcommand{\thetable}{S\arabic{table}}
\def\theequation{S\arabic{equation}}

\setcounter{figure}{0}
\setcounter{table}{0}
\setcounter{equation}{0}

\section{Musculoskeletal Hand Model}

Accurately modeling the human hand remains a challenging problem due to its highly complex musculoskeletal structure, which involves dozens of muscle-tendon units with subtle and coordinated functions. 
While MyoHand~\cite{caggiano2022myosuite} provides an anatomically grounded model of muscle-tendon structures and joint configurations based on previous work in biological modeling~\cite{mcfarland2019spatial,lee2015finger}.
We augment the original MyoHand model with five intrinsic muscle-tendon units located in the thumb-side and little-finger-side regions of the palm: FPB (flexor pollicis brevis), APB (abductor pollicis brevis), AdP (adductor pollicis), FDM (flexor digiti minimi brevis), and ADM (abductor digiti minimi), as shown in Fig.~\ref*{fig:hand_anatomy}. These muscles are selected to enhance the abduction-adduction and flexion of the thumb and the pinky finger, allowing their independent control. This combination of flexion and lateral spreading is critical for wide-span configurations and dexterous finger coordination in piano performance.

Each added muscle-tendon unit is implemented using the same abstraction as in MyoHand. Specifically, each muscle is modeled as an independent actuator with a single activation variable, simplified straight-line routing, and via points where necessary to approximate physiological moment arms. These attachments are added to the existing skeletal model without altering the original joint definitions. The added intrinsic actuators share the same activation dynamics, force-length-velocity relationships, and parameter scaling rules as the original MyoHand muscles, ensuring consistency and reproducibility within the existing simulation framework. Key muscle properties, such as peak force and optimal fiber length, were derived from prior literature~\cite{holzbaur2005model,del2007relation}, and scaled following the conventions used for other muscles in the MyoHand model, while muscle attachment sites were defined based on standard anatomical atlases~\cite{rohen2006color,standring2005gray}.
The impact of our newly added muscle-tendon units on dexterous control is highlighted in the experiment section. We refer to the supplementary video for animated results.

Besides, following MyoHand, to balance anatomical fidelity and computational tractability, our model adopts a simplified representation that focuses on task-relevant structures and motions. As a result, certain degrees of freedom associated with palmar bones and palm-level articulations, whose motions are typically small, are not explicitly modeled.

\begin{figure}
    \centering
    \includegraphics[width=\linewidth]{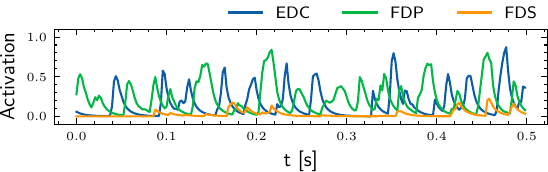}

    \vspace{0.5cm}
    
    \includegraphics[width=.7\linewidth]{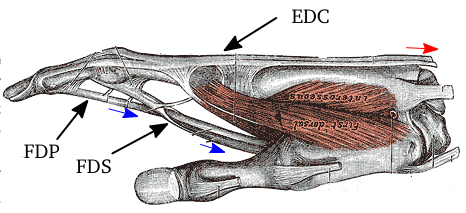}
    \caption{Top: Activation patterns generated by the agonist-antagonist muscle pair of EDC, FDP, and FDS during playing \textit{Rhapsodie Op.79 No.2}. The functions of EDC, FDP and FDS for finger control are illustrated in the subfigure at the bottom. While muscles can only generate unidirectional forces shown by the arrows in red and blue, the finger movement is decided by the relative force generated by these muscles. Our model can generate sparse activations with staggered peaks, which reflect an energy-efficient, human-like control strategy for agonist and antagonist muscles.}
    \label{fig:agonist}
\end{figure}

In real biological systems, muscles can generate force only through contraction (pulling) and cannot actively produce force in the opposite direction (pushing).
This characteristic is reflected during simulation by only allowing a musculotendon actuator to generate unidirectional force.
Consequently, movements that require bidirectional control - such as flexion and extension - have to be achieved by using pairs of muscles with opposing mechanical actions, commonly referred to as agonist and antagonist muscle groups. Because human motor control is generally optimized for energy efficiency, agonist and antagonist muscles are typically not strongly activated simultaneously, resulting in a relatively sparse and alternating activation pattern~\cite{tresch2009case,grillner1985neurobiological,sherrington1910flexion,ting2007neuromechanics}.
Such behavior is reflected in the control strategy learned by our reinforcement learning policy. 
As shown in Fig.~\ref{fig:agonist}, a clear antagonistic relationship emerges between the flexor muscles (FDP and FDS) and the extensor muscle (EDC). Specifically, when FDP and FDS exhibit large activation, EDC activity remains at a relatively low level, and vice versa. 
The resulting muscle coordination closely resembles physiological motor control, in which agonist and antagonist muscles are activated in an energy-efficient manner. 
Notably, these results are generated by our high-level control policy, which is isolated from the musculotendon states and performs control by operating the latent.
This indicates that the learned low-level policy captures a biologically plausible control strategy while serving as a stable low-level muscle-driven controller.

\begin{figure}
    \centering
    \includegraphics[width=\linewidth]{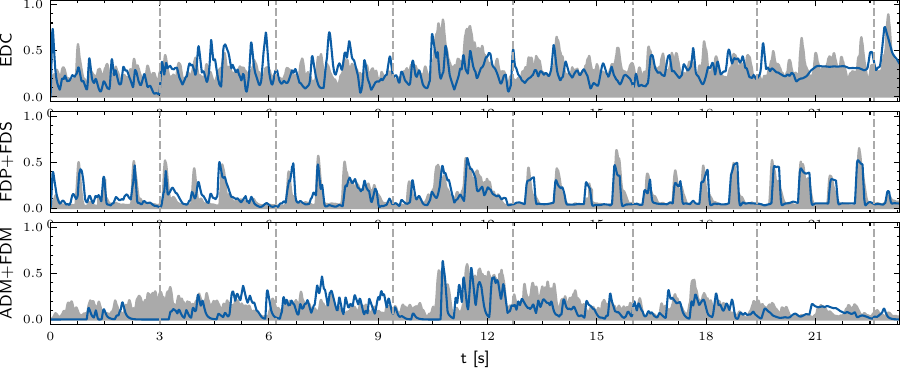}
    \caption{Comparison of EMG recordings collected from a human subject (shadow regions) during the performance of \textit{Mary Had a Little Lamb}, and the activations (blue lines) generated by our model while tracking the motions.}
    \label{fig:emg_mary}
\end{figure}

To further evaluate the physiological fidelity of our model, we perform an additional comparison in Fig.~\ref{fig:emg_mary}, where EMG recordings were collected while a subject performed the simple piece \textit{Mary Had a Little Lamb}, consisting of 26 key strokes. We refer to Sec.~\ref*{sec:exp_hand_model} for details on sensor placement. While noticeable noise is present in the EDC and ADM+FDM channels, due to wrist motion and the inherent limitations of surface EMG in resolving individual muscle contributions, the generated activations still broadly align with the recorded signals. In the FDP+FDS channel, the synthesized activations closely capture the key stroke events, exhibiting 26 distinct peaks that correspond well with the EMG measurements. These results suggest that our model is capable of producing physiologically consistent activation patterns that reflect the underlying structure of human muscle activity during piano performance.

\section{Data Preprocessing}

Our reference motion data are extracted from  F\"urElise~\cite{wang2024furelise}.
We retarget the original motions to our musculoskeletal hand models,
and filter out segments with large jittering and invalid joint rotations and those only containing still hand motions.
The resulting dataset contains 651 minutes of two-hand motion data. 

While the source dataset provides annotations from the wrist to fingers, it lacks whole-body context, omitting elbow and shoulder positioning as well as wrist abduction and flexion. 
However, our hand model has the forearm,
as many muscles responsible for finger and wrist motion originate there.
Our retargeting recovers elbow motion by constraining it to the approximate height of the keyboard while penalizing excessive shoulder displacement.

\begin{table}
    \caption{Joint parameters used during simulation. The hand joint parameters follow MyoHand~\cite{caggiano2022myosuite}, while the piano key parameters are adopted from RoboPianist~\cite{zakka2023robopianist}. We refer to F\"urElise~\cite{wang2024furelise} for the geometric dimensions of the piano.}
    \centering
    \begin{tabular}{lcc}
    \toprule
    \multicolumn{1}{c}{} & \multicolumn{1}{c}{Hand Joint} & Piano Key \\
    \midrule
        damping & $0.05$ & $0.05$ \\
        stiffness & $0$ & $2$\\
        armature & $0.0001$ & $0.001$\\
        solref & $0.02\ 1$ & $0.01\ 1$ \\
        solimp & $0.8\ 0.8\ 0.01$ & $0.95\ 0.99\ 0.001$ \\
    \bottomrule
    \end{tabular}
    \label{tab:phys_parameters}
\end{table}

\begin{table}
\centering
\tiny
\caption{Implementation details of motion retargeting.}
\begin{tabular}{l|cccc}
    \toprule
Stage & Iters. & LR & Loss & Param.  \\
\midrule
I & 1000 & $0.01$ & $\mathcal{L}_{\text{recon}} + 0.01 \mathcal{L}_{\text{smooth}} $ & $\mathbf{q}_{\text{hand}}$ \\
II & 1000 & $0.01$ & $\mathcal{L}_{\text{recon}} + 0.01 \mathcal{L}_{\text{smooth}} + \mathcal{L}_{\text{elbow}} $ & $\mathbf{q}_{\text{elbow}}$ \\
III & 1000 & $0.001$ &  $\mathcal{L}_{\text{tip}} + \mathcal{L}_{\text{delta}} + 10\mathcal{L}_{\text{violation}} $ & $\mathbf{q}_{\text{hand}}, \mathbf{q}_{\text{elbow}}$ \\
IV & 2000 & $0.001$ & $\mathcal{L}_{\text{tip}} + 0.01 \mathcal{L}_{\text{smooth}} + \mathcal{L}_{\text{delta}} + 10\mathcal{L}_{\text{violation}} $ & $\mathbf{q}_{\text{hand}}, \mathbf{q}_{\text{elbow}}$ \\
\bottomrule
\end{tabular}
\label{tab:ik}
\end{table}

\begin{table}
    \centering
    \caption{Definitions of the loss terms for motion retargeting.}
    \begin{tabular}{l|l}
    \toprule
        $\mathcal{L}_{\text{recon}}$ & $\sum_{t} \sum_{j\in \mathcal{M}_{all}}\|\mathbf{x}_t^j - FK(\mathbf q_t) \| ^ 2$ \\
        $\mathcal{L}_{\text{tip}}$ & $\sum_{t} \sum_{j\in \mathcal{M}_{tip}}\|\mathbf{x}_t^j - FK(\mathbf q_t) \| ^ 2$ \\
        $\mathcal{L}_{\text{smooth}}$ & $\sum_{t} \|\ddot {FK(\mathbf q_t)} \| ^ 2$ \\
        $\mathcal{L}_{\text{delta}}$ & $\sum_{t} \|\mathbf{q}_t - \hat{\mathbf{q}}_t \| ^ 2$ \\
        $\mathcal{L}_{\text{viol}}$ & $\sum_{t} \left( \| [\mathbf{q}_{\text{min}} - \mathbf{q}_t]_+ \|_1 + \| [\mathbf{q}_t - \mathbf{q}_{\text{max}}]_+ \|_1 \right)$ \\
        $\mathcal{L}_{\text{elbow}}$ & $\sum_{t} ( z_t^{\text{elbow}} - h_{\text{piano}} )^2$ \\
    \bottomrule
    \end{tabular}
    \label{tab:ik_loss}
\end{table}

The retargeting pipeline employs a four-stage optimization to produce natural motion from the elbow to fingertips. Implementation details, including optimized parameters, iteration number, learning rates (LR), and loss terms, are detailed in Table~\ref{tab:ik}.
The detailed definitions of the loss terms are presented in Table~\ref{tab:ik_loss}.
where $FK(\cdot)$ denotes a differentiable operator of forward kinematics of the hand model, $\mathbf{x}_t^j$ represents the annotated position of joint $j$ in the original dataset, and $\hat{\mathbf{q}}$ corresponds to the initial configuration before the current optimization stage, while $\mathbf{q}_{\min}$ and $\mathbf{q}_{\max}$ define the joint limits.

\section{Implementation Details}
We provide additional implementation details regarding policy training and adaptive sampling in this section.

\begin{table}
\centering
\caption{Hyperparameters}
\begin{tabular}{llc}
    \toprule
     & \textbf{Parameter} & \textbf{Value}\\
    \midrule
    \multicolumn{2}{l}{\textit{General Reinforcement Learning Settings}} & \\
     & policy network learning rate & $5 \times 10^{-6}$\\
     & critic network learning rate & $1 \times 10^{-4}$\\
     & reward discount factor ($\gamma$) & $0.95$ \\
     & GAE discount factor ($\lambda$) & $0.95$ \\
     & surrogate clip range ($\epsilon$) & $0.2$ \\
     & PPO batch size & $256$ \\
     & PPO optimization epochs & $5$ \\
    \midrule
    \multicolumn{2}{l}{\textit{Tracking Policy Training}} & \\
     & simulation instances (number of environments) & $8192$ \\
     & rollout length & $32$ \\
     & tracking target horizon ($H$) & $4$ \\
     & adaptive sampling discount factor ($\zeta$) & 0.99 \\
     & adaptive sampling power scale factor ($\eta$) & 5 \\
     & adaptive sampling moving average coefficient ($\alpha$) & 0.5 \\
     & episode length ($C$) & 1440 \\
     & latent dimension ($\dim \mathbf{h}_t$) & 32 \\
    \midrule
    \multicolumn{2}{l}{\textit{VAE Training}} & \\
    & learning rate & $3 \times 10^{-4}$\\
    & KL weighting coefficient ($\beta$) & $0.005$ \\
    & batch size & $4096$ \\
    & simulation instances & $4096$ \\
     & latent dimension ($\dim \mathbf{z}_t$) & 32 \\
    \midrule
    \multicolumn{2}{l}{\textit{High-Level Controller Training}} & \\
     & simulation instances (number of environments) & $2048$ \\
     & rollout length & $8$ \\
     & target key horizon ($N$) & $5$ \\
     & adaptive sampling discount factor ($\zeta$) & 0.95 \\
     & adaptive sampling power scale factor ($\eta$) & 8 \\
     & adaptive sampling moving average coefficient ($\alpha$) & 0.5 \\
     & episode length ($C$) & 150 \\
     & discriminator learning rate & $1 \times 10^{-5}$\\
     & discriminator batch size & $512$ \\
     & gradient penalty coefficient ($\lambda^{GP}$) & $10$ \\
     & goal-directed objective weight for key press & $0.8$ \\
     & imitation objective weight (wrist and elbow)  & $0.05$ \\
     & imitation objective weight (fingers)  & $0.15$ \\
  \bottomrule
\end{tabular}
\label{tab:hyper}
\end{table}

\begin{table}
\centering
\caption{Weights of each body link, named by the corresponding bone, for tracking error computation. For each finger, we consider the three associated parts of \textcolor{violet}{proximal}, \textcolor{blue}{middle}, and \textcolor{green}{distal} phalanges, highlighted in purple, blue, and green, respectively. Specially, the thumb has a controllable \textcolor{orange}{metacarpal} (orange), while it does not have the middle phalanx. Therefore, we use the metacarpal, and proximal and distal phalanges for the thumb instead. The wrist position is decided by the \textcolor{red}{lunate} bone, which is highlighted in red. The elbow, which is a free joint with 6 DoFs, is located at the ulna. The other bones inside the palm are modeled as immovable. All weights will be normalized by L-1 before used for reward computation.}

\includegraphics[width=0.85\linewidth]{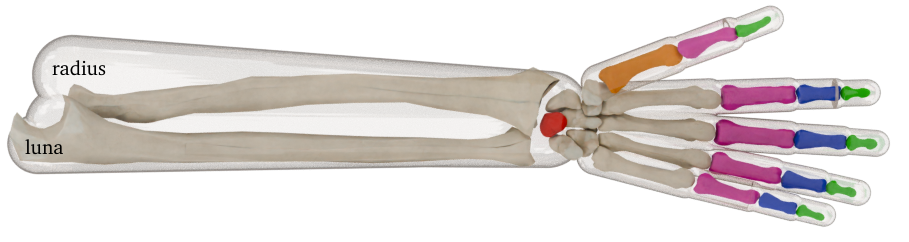}
\begin{tabular}{l|c|c}
    \toprule
     Bone Name & $\omega_i$ & $\kappa_i$ \\
     \midrule
     ulna & 0.1 & 0.1\\
     radius & 0.05 & 0\\
     \textcolor{red}{lunate} & 0.2 & 0.1\\
     \textcolor{violet}{proximal phalanx}  & 0.1 & 0\\
     \textcolor{blue}{middle phalanx} & 0.1 & 0\\
     \textcolor{green}{distal phalanx} & 0.1 & 0 \\
     \textcolor{orange}{metacarpal} (thumb) & 0.1 & 0 \\
     \textcolor{violet}{proximal phalanx} (thumb) & 0.1 & 0\\
     \textcolor{green}{distal phalanx} (thumb) & 0.1 & 0.1 \\
     \textcolor{green}{distal phalanx}  (pinky) & 0.1 & 0.1 \\
     fingertips & 0 & 0.2 \\
     \bottomrule
\end{tabular}
\label{tab:track_weights}
\end{table}

\subsection{Tracking Policy Training}
The hyperparameters used during tracking policy training are listed in Table~\ref{tab:hyper}.
In Table~\ref{tab:track_weights}, we summarize the weights for position and orientation error computation used in the reward function (c.f. Eq.~\ref*{eq:tracking_reward}).
All weights will be normalized such that $\sum_i\omega_i = 1$ and $\sum_i\kappa_i = 1$ before incorporated in the reward function.
The position error is computed only for the elbow, wrist and end-effector fingertips, except for the thumb and pinky, which additionally take into account the distal phalanx to improve their flexion performance.
We use a multilayer perceptron with hidden neurons of 1024, 1024, and 512 at each layer in the target pose encoder $\mathcal{G}$, while the same architecture is adopted by the tracking policy $\pi_\text{track}$ and the VAE encoder and decoder.
The hidden latent $\mathbf{h}_t$ is fed into $\pi_\text{track}$ by concatenating with the proprioception state $\mathbf{s}_t$. 
We use a standard implementation of PPO-Clip~\cite{schulman2017proximal} (no clipping on the state value estimation) to perform training.
All networks are updated using Adam optimizers~\cite{kingma2014adam}.

\subsection{Latent Distillation}
We adopt on-policy distillation in a hybrid manner. 
Specifically, rather than relying exclusively on the actions generated by the VAE model to actuate the hand during online simulation, we advance the simulation, with a 20\% chance, using actions from the tracking policy.
This hybrid scheme stabilizes training by preventing frequent simulation failures caused by stochastic or poorly generalized VAE actions in early stages of training, while still allowing the VAE to learn under realistic on-policy state distributions.
An early termination will be triggered if the tracking error is too large (greater than $0.5$m) during distillation.
The pseudocode of the distillation process is provided in Algorithm~\ref{alg:distill}.

\begin{algorithm}[t]
\SetKwInOut{Input}{Input}
\SetKwInOut{Output}{Output}
\nl Initialize $N$ environments running in parallel where $N$ is the batch size for network updating; \\
\nl \While{training does not converge}{
\nl $\mathbf{s}_t, \mathbf{q}_t^\text{target} \leftarrow$ env.state(), env.tracking\_target();\\
\nl $\mathbf{z}_t \sim q_\theta(\cdot|\mathbf{s}_t,\mathbf{h}_t)$ where $\mathbf{h}_t = \mathcal{G}(\mathbf{q}_t^\text{target})$;\\
\nl \For{each substep $\delta$}{
\nl $\mathbf{\hat{a}}_{t+\delta} = \textsc{Dec}(\mathbf{s}_{t+\delta},\mathbf{z}_t)$; \\
\nl $\mathbf{\bar{a}}_{t+\delta} = \textsc{Clip}(\mathbf{a}_{t+\delta})$ where $\mathbf{a}_{t+\delta}\sim\pi_\text{track}(\cdot|\mathbf{s}_{t+\delta}, \mathbf{h}_t)$; \\
\nl Update $\textsc{Dec}$ and $q_\theta$ using Eq.~\ref*{eq:vae}; \\
\nl \uIf{rand() < 0.2}{
\tcp{Step environment using actions from $\pi_\text{track}$.}
\nl $\mathbf{s}_{t+\delta+1}\leftarrow$  env.step($\mathbf{\bar{a}}_{t+\delta}$).
}
\nl\uElse{
\tcp{Step environment using actions from VAE.}
\nl $\mathbf{s}_{t+\delta+1}\leftarrow$  env.step($\mathbf{\hat{a}}_{t+\delta}$).
}
}
}
\caption{Hybrid On-Policy VAE Distillation}
\label{alg:distill}
\end{algorithm}

\subsection{High-Level Motion Synthesis}\label{sec:high_appendix}
\begin{figure}
    \centering
    \includegraphics[width=\linewidth]{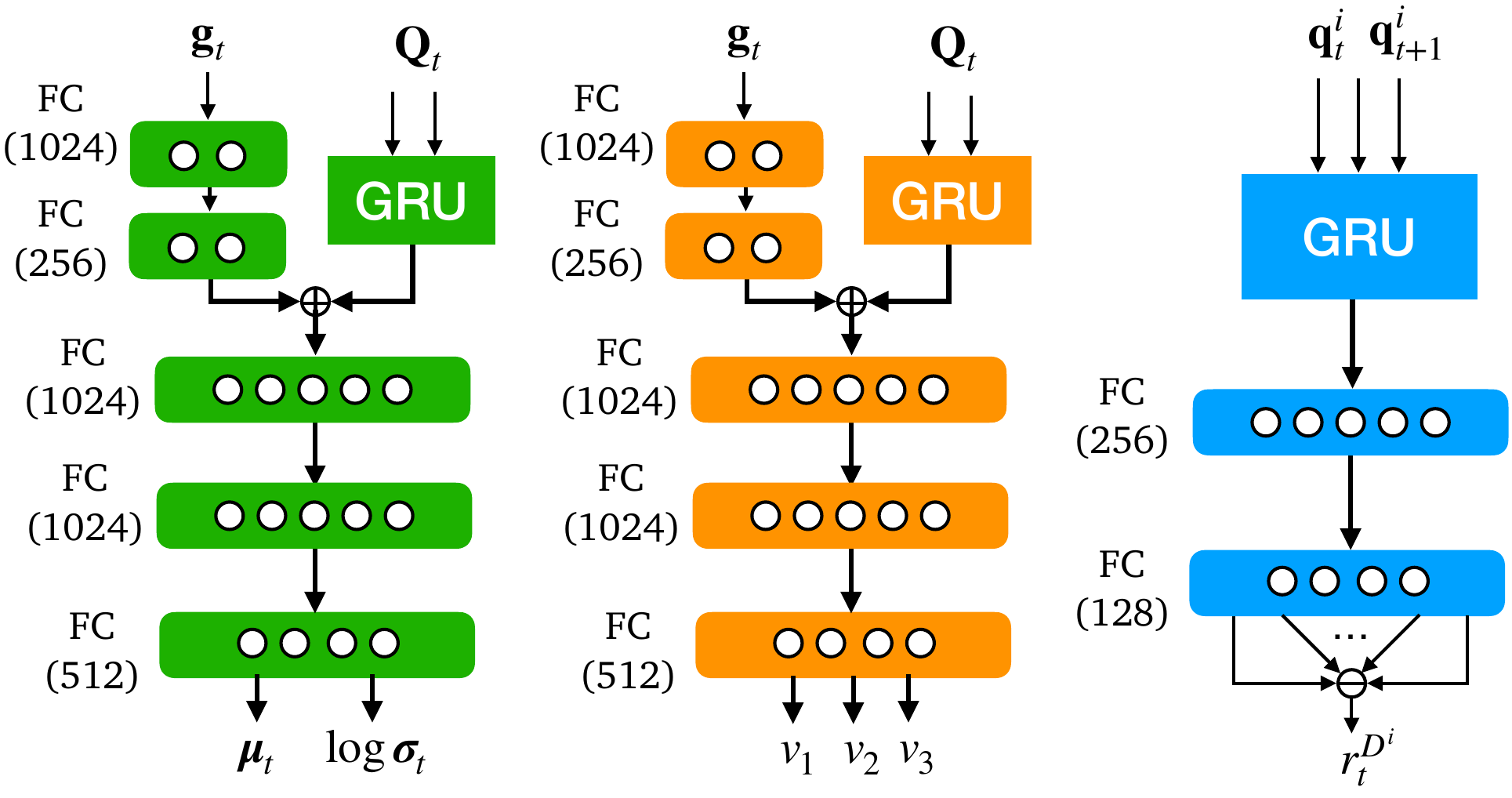}
    \caption{Network structures used by the high-level motion synthesis policy. From left to right, the policy network, value network, and discriminator network. We use $\oplus$ denoting the add operator and $\ominus$ denoting the average operator. The value network employs a multi-head architecture to estimate the state values associated with the goal-directed objective and two imitation objectives separately. A similar architecture is applied to the discriminator network, which serves as a discriminator ensemble by providing a multi-dimensional output.}
    \label{fig:network}
\end{figure}

Similar to the tracking policy, the high-level controller for motion synthesis is trained using PPO~\cite{schulman2017proximal} as the backbone reinforcement learning algorithm. 
The hyperparameters used for policy training are listed in Table~\ref{tab:hyper} and the network structures are illustrated in Fig.~\ref{fig:network}.
We take a GAN-like architecture~\cite{iccgan} to incorporate imitation rewards. 
Instead of directly imitating whole hand motions,  
we use two discriminators for each hand to perform motion imitation: one monitors the global positioning of the wrist and elbow relative to the piano, while the other focuses on the local pose of the fingers relative to the wrist.
In such a way, the policy is allowed to freely combine finger poses with various wrist placements.
This imitation learning setup, plus the goal-directed reward (Eq.~\ref*{eq:goal_reward}), results in three optimization objectives. 
We perform multi-objective learning using the framework from prior literature~\cite{xu2023composite},
and employ a multi-head critic network to estimate the advantage associated with each objective.
The optimization function of reinforcement learning can be written as
\begin{equation}
    \max \mathbb{E}\left[\sum_k w_k \bar{A}_{t,k} \log \pi_\text{high}^h (\mathbf{z}_t | \mathbf{Q}_t, \mathbf{g}_t)\right]
\end{equation}
where $\bar{A}_{t,K}$ is the standardized advantage that is estimated according to the reward associated with each objective $k$, $w_k$ is the associated weights (see Table~\ref{tab:hyper}), and $h \in \{\text{left}, \text{right}\}$ indicating the target hand.
The imitation reward provided by the discriminator is obtained by
\begin{equation}
    r_t^\text{imit}(\mathbf{q}_t^i, \mathbf{q}_{t+1}^i) = \frac{1}{M} \sum_{m=1}^M \textsc{Clip}\left(D_m^i(\mathbf{q}_t^i, \mathbf{q}_{t+1}^i), -1, 1\right)
\end{equation}
where the subscript $i$ indicates the different imitation objective.
While the policy input $\mathbf{Q}_t$ is the set of each body link's pose (position and orientation in Cartesian space) and linear and angular velocities of the two hands with a 2-frame historical horizon,
the input to the discriminator ($\mathbf{q}_t^i$ and $\mathbf{q}_{t+1}^i$) denotes the partial observation of the link poses associated with the imitation objective $i$ and the policy observation $\mathbf{Q}_t$ and $\mathbf{Q}_{t+1}$, i.e., the partial link poses at the timesteps from $t-1$ to $t+1$.
Here, we take an ensemble of $M=32$ discriminators to evaluate imitation performance.
The discriminator is trained using hinge loss~\cite{lim2017geometric} with gradient penalty~\cite{gulrajani2017improved}.
We refer to the previous literature~\cite{iccgan} for details of discriminator training.

\subsection{Adaptive Sampling}
To improve the efficiency of adaptive sampling,
the interval between chunks is expected to be small.
Such that we can have more densely distributed chunks, enabling precise localization of segments where tracking performance degrades.
However, the increased number of chunks requires a longer sampling period to update the performance estimates of all chunks.
In our approach, the number of chunks, which determines the chunk overlapping given that all chunks are equally long, is chosen based on the following formula:
\begin{equation}
    \#\{\text{epochs}\} \approx \frac{\#\{\text{chunks}\}}{\#\{\text{environments}\}}\times\frac{\text{chunk length}}{\text{rollout length}}.
\end{equation}
This formula approximates the number of training epochs required to update performance estimates of all chunks when multiple simulation environments run in parallel and sampling weights are approximately uniform.
In our implementation, we take $20$ as the desired number of updating epochs to decide the number of chunks.
For high-level policy training, we simply perform chunking at the note level.

\begin{figure}
    \centering
    \includegraphics[width=\linewidth]{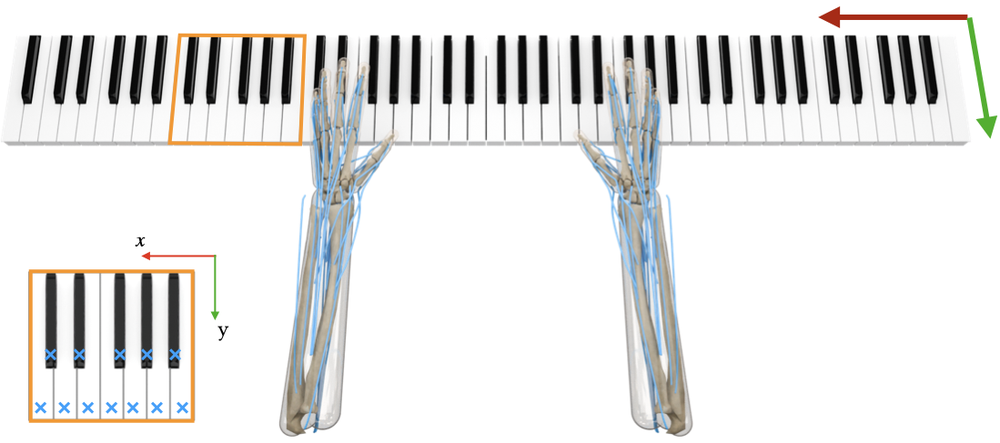}
    \caption{Reference axis and position definitions of piano keys. The key positions $\mathbf{p}_k$ used during reward computation are defined on the top surface of the key, and marked by cross symbols in the window at the bottom-left. At the beginning of each training episode, the hands will be initialized above the keyboard via a fixed pose to avoid penetration with the keyboard model.}
    \label{fig:piano_axis}
\end{figure}
The estimated policy performance is updated online during policy training through a moving average:
\begin{equation}
    \bar{r}_i \leftarrow (1-\alpha) \bar{r}_i + \alpha \bar{r}_i^\prime
\end{equation}
where the coefficient $\alpha$ should be large enough to avoid sluggish updates.
We take $\alpha=0.5$ in our implementation for both the tracking policy and the high-level motion synthesis policy training.

\subsection{Piano Model}
The piano model follows a standard design, with an octave span of 16.5 cm (6.5 in). Our hand model can span approximately eight white keys between the thumb and pinky, corresponding to the average hand size reported for female pianists~\cite{boyle2015pianist}. However, this span is smaller than that required by some piano repertoire. Consequently, our experiments focus on repertoire that does not require a larger hand span.

\begin{figure}
    \centering
     \includegraphics[width=\linewidth]{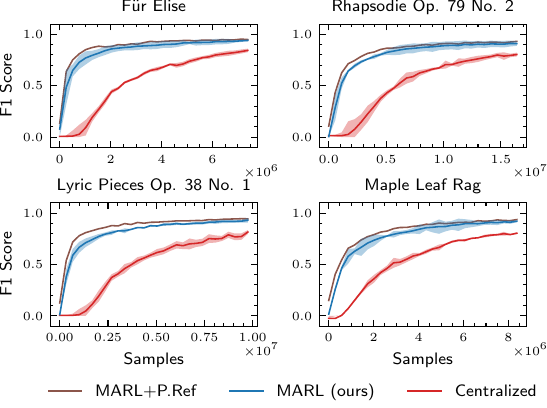}
    \caption{Learning performance of the high-level controllers using a multi-agent setup (MARL) and that using a centralized policy for bimanual control (Centralized). MARAL+P.Ref indicates the baseline using our multi-agent setup with prior reference motions provided by a joint-driven policy (Sec.~\ref{sec:prior_ref}.) The Centralized setup typically needs around 30\% more training iterations to reach the same final performance as MARL.}
    \label{fig:ablation_marl_prior}
\end{figure}

\begin{figure}
    \centering
    \includegraphics[width=0.75\linewidth]{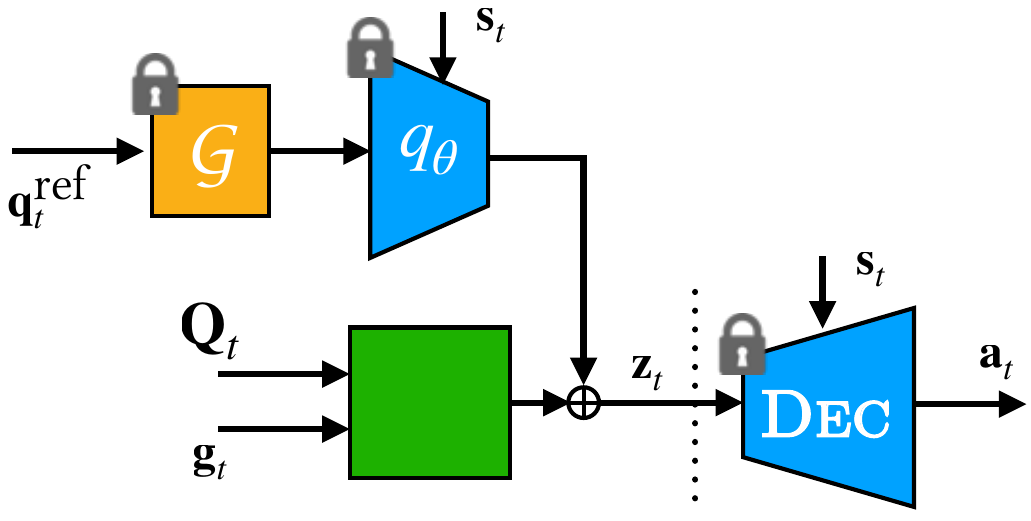}
    \caption{Diagram to incorporate prior reference motions to facilitate high-level policy training. $\oplus$ denotes the add operation. In Fig.~\ref{fig:ablation_marl}, we show a baseline using a joint-driven control policy to provide the prior reference $q_t^\text{ref}$. Similar approaches can be applied while using reference motions generated by other models, e.g., a diffusion model~\cite{wang2024furelise}.}
    \label{fig:net_prior_ref}
\end{figure}
To improve simulation speed, we modify the underlying collision detection implementation of MuJoCo-MJX, and adopt a dynamic collision configuration for piano keys.
Specifically, before each simulation step, we compute the bounding box defined by the fingertips and wrist, and enable collisions only for the 18 keys (including both white and black keys) directly below the bounding box,
while one octave has seven white keys and five black keys.
This configuration will bring about an improvement of around 20\% on simulation speed, and largely reduce the GPU memory needed by the simulator to process contact.

In the reward term presented in Eq.~\ref*{eq:r_k_pos}, the distance between a target key $k$ and the associated finger $i$ is defined as $\mathbf{d}_{k,i} = \mathbf{p}_k - \mathbf{p}_i$, where $\mathbf{p}_i$ is the position of the fingertip $i$ and $\mathbf{p}_k$ is the position of the key $k$ in the Cartesian space.
Given the bounding box of a key defined via the lower bound $(x_0, y_0, z_0)$ and upper bound $(x_1, y_1, z_1)$, we define $\mathbf{p}_k := \left(0.5(x_0+x_1), y_0+0.9(y_1-y_0), z_1\right)$.
Referring to the axis definition illustrated in Fig.~\ref{fig:piano_axis},
the key position corresponds to the lateral center on the top surface of each key and at the 90\%-th position longitudinally towards the performer.
Specially, to allow a flexible approach along the longitudinal side, the $y$-axis distance is downscaled by 90\% when the finger is above the keyboard, i.e., $d_{k,i}^y = 0.1 (p_k^y - p_i^y)$ if finger $i$ is over the keyboard.
The $z$-axis distance is ignored if the finger is already pressing the target key, i.e., $d_{k,i}^z = 0$ if $p_i^z < p_k^z$ and $\mathbb{I}_{k,i} = 1$, 
where $\mathbb{I}_{k,i}$ is an indicator evaluating if finger~$i$ lies over the key $k$ by checking the fingertip's position and the key's bounding box on the $xy$-plane.

\begin{figure}
    \centering
    \includegraphics[width=0.325\linewidth]{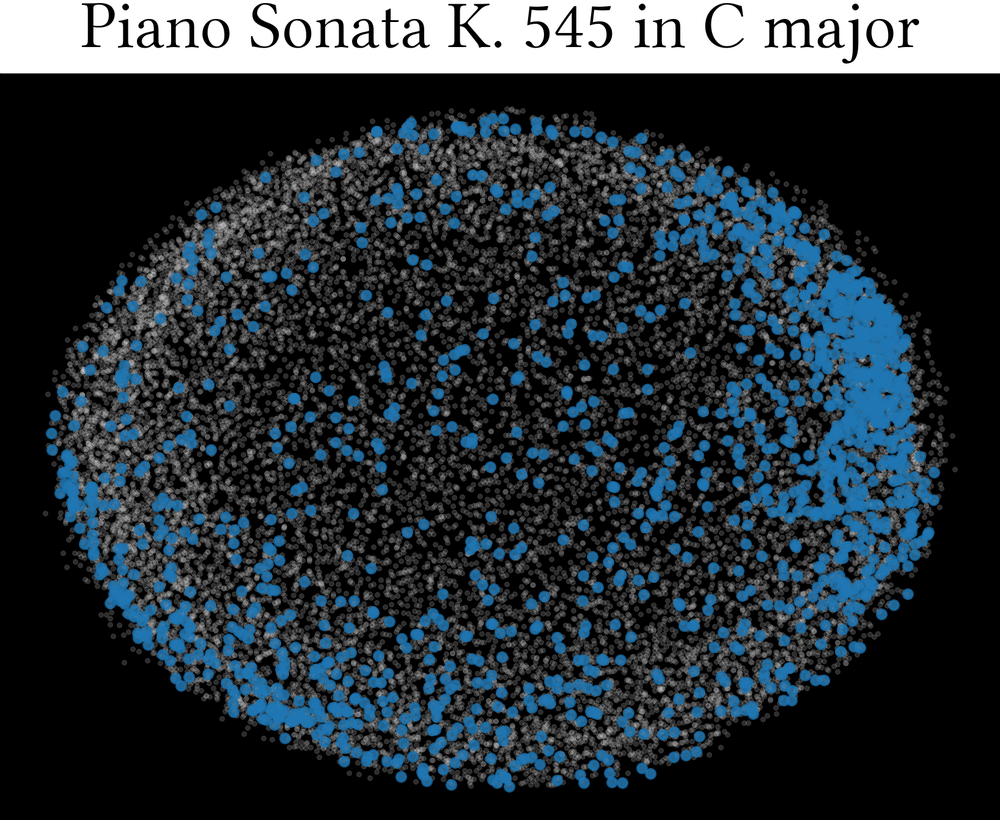}
    \includegraphics[width=0.325\linewidth]{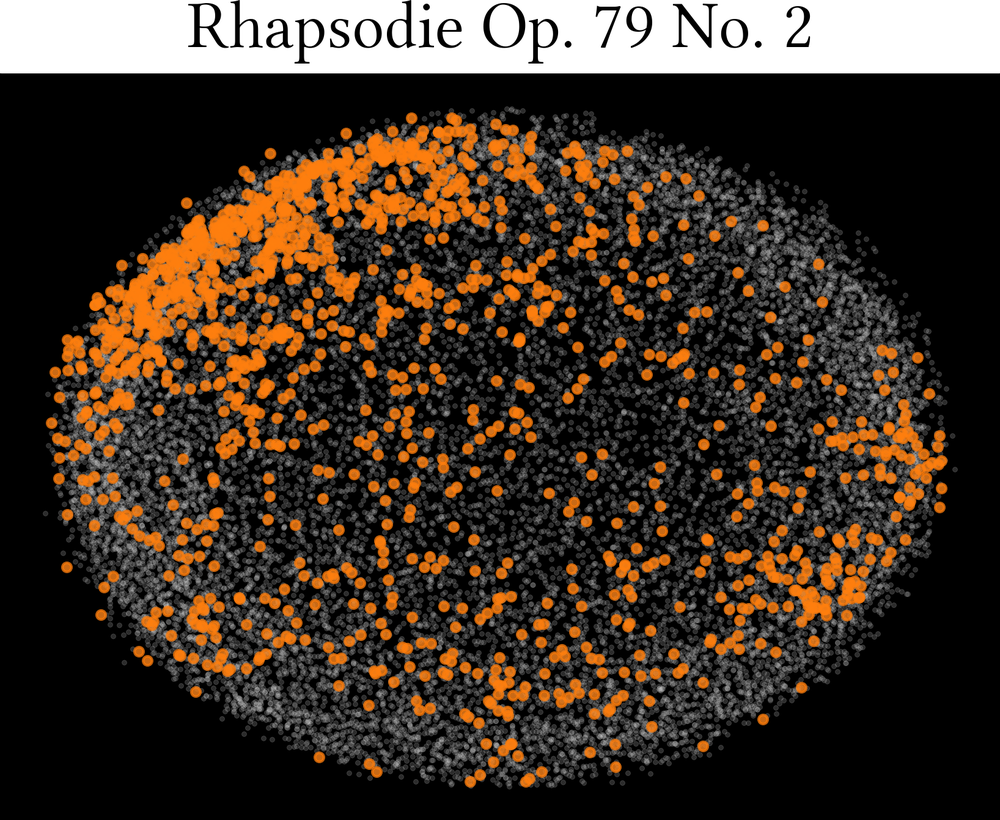}
    \includegraphics[width=0.325\linewidth]{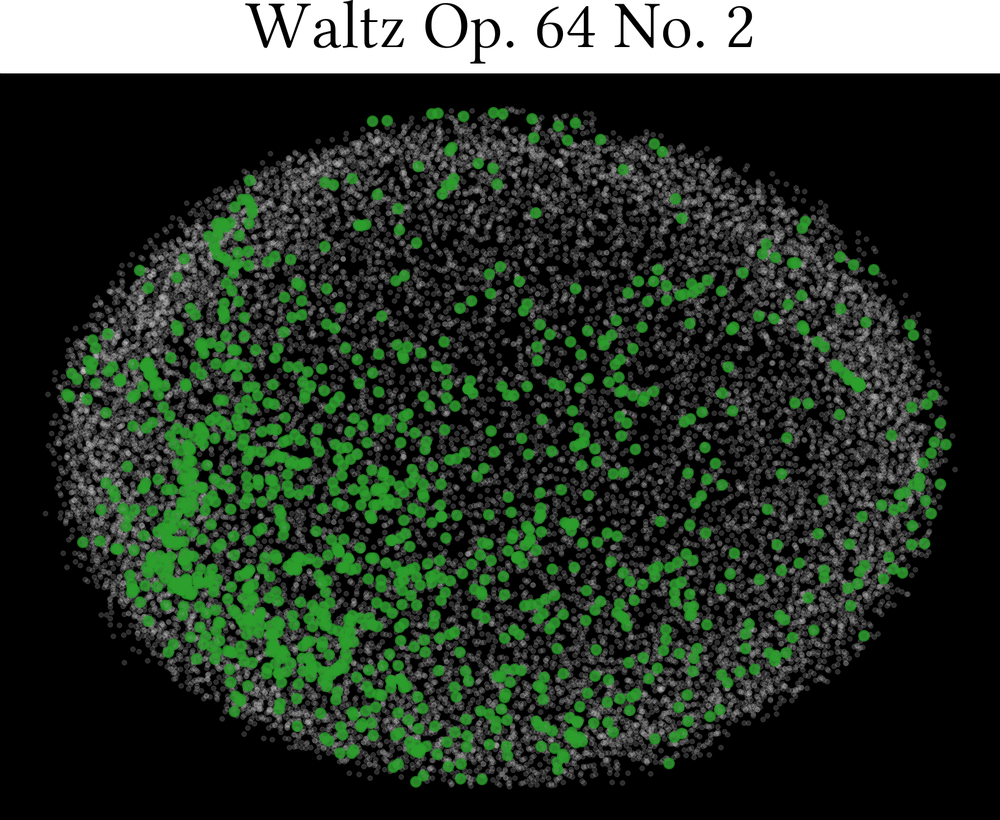}\\
    \vspace{.5em}
    \includegraphics[width=0.3259\linewidth]{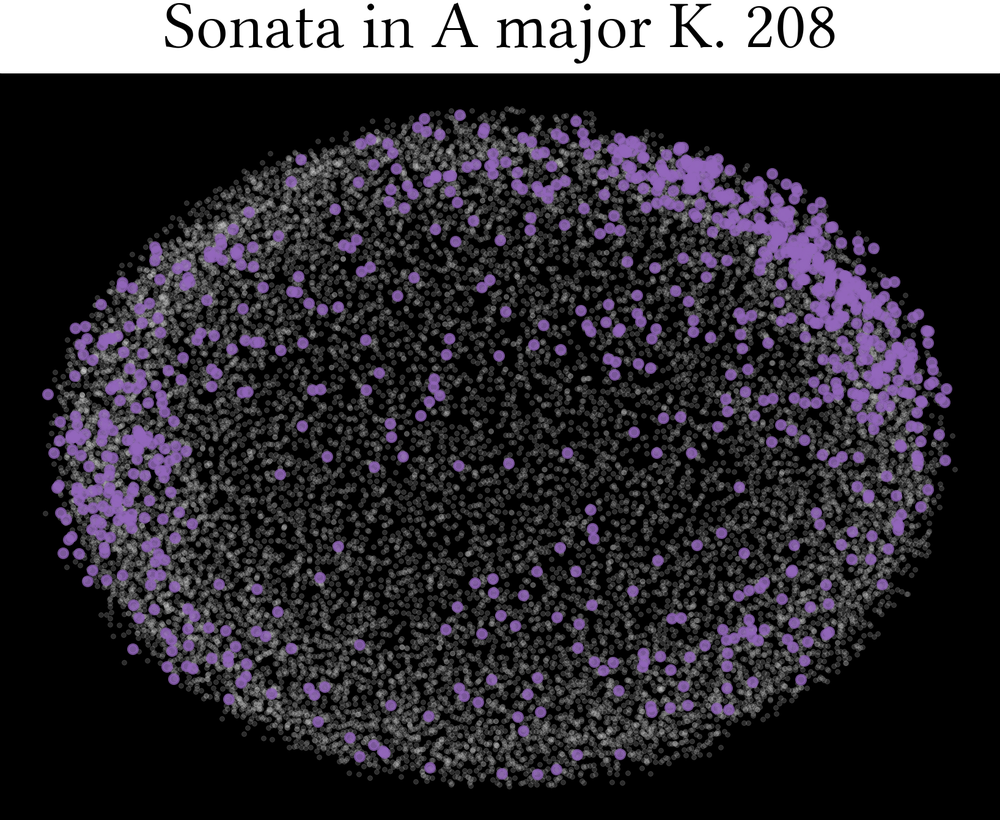}
    \includegraphics[width=0.3259\linewidth]{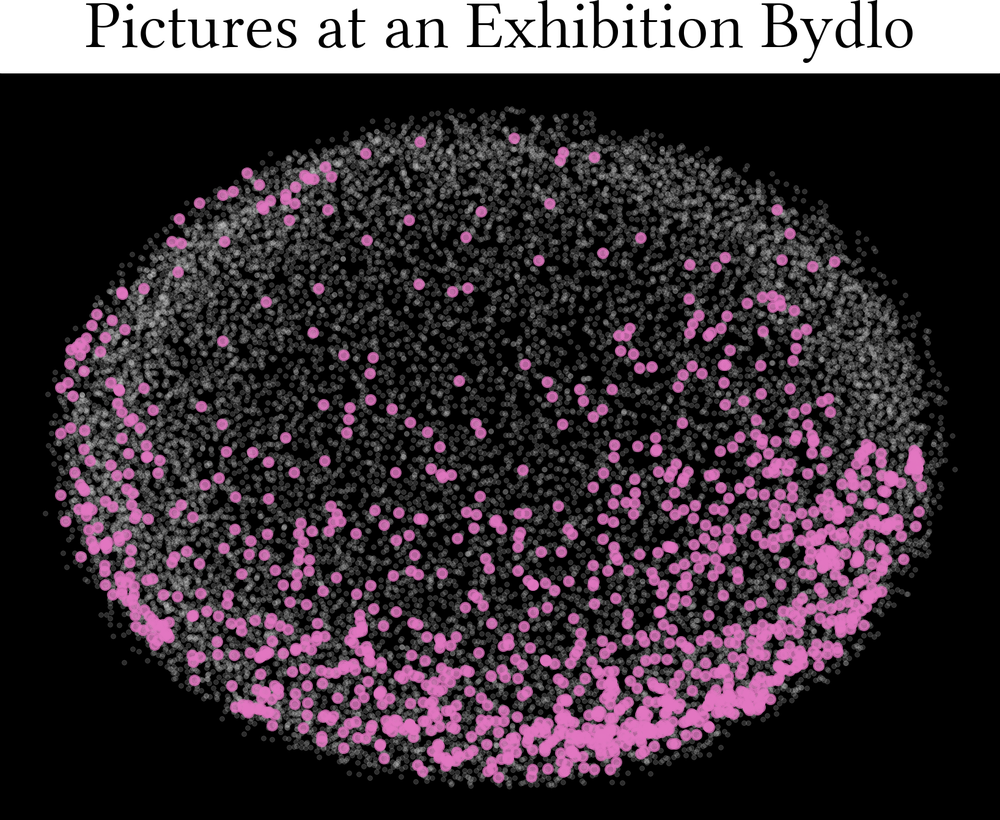}
    \includegraphics[width=0.3259\linewidth]{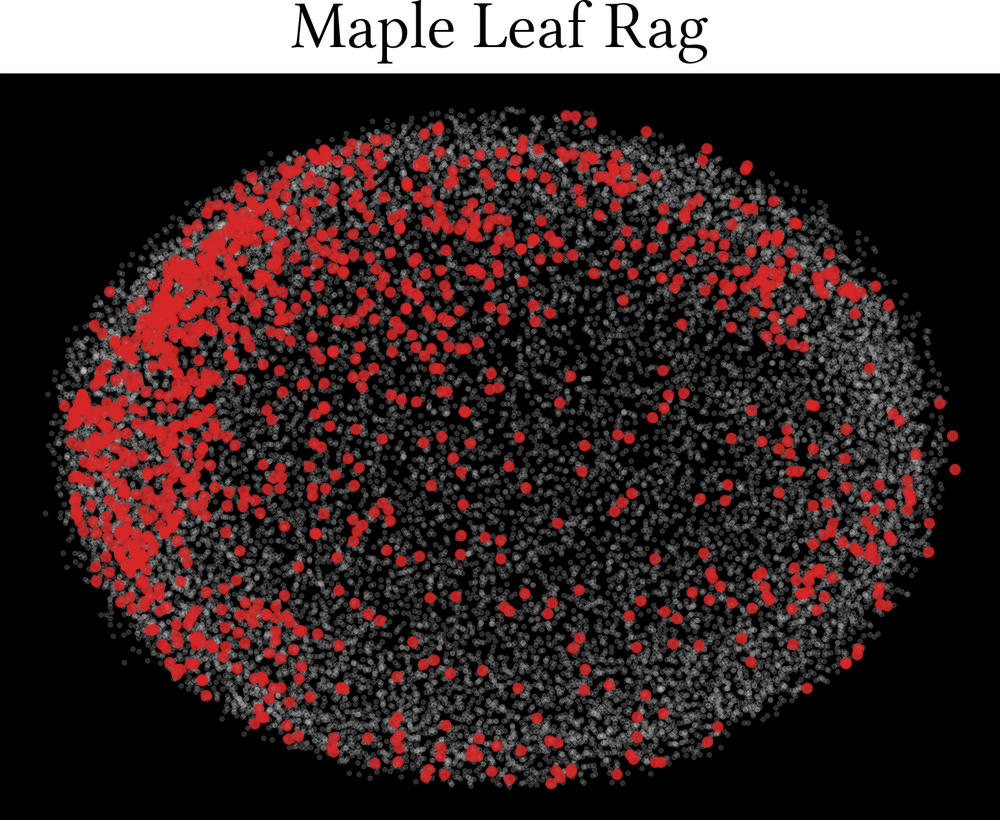}
    \caption{Visualization of the VAE latent used by the high-level policies to perform motion synthesis.
    The gray dots denote the latent used by the high-level policies across all our tested musical pieces, with the highlighted ones indicating those used for the specific piece of music.
    The visualization is achieved using the multidimensional scaling technique to project the latent of two hands from $32\times2$ dimensions to 2 dimensions.
    }
    \label{fig:latent}
\end{figure}

\begin{figure*}[htp]
    \centering
    \includegraphics[width=\linewidth]{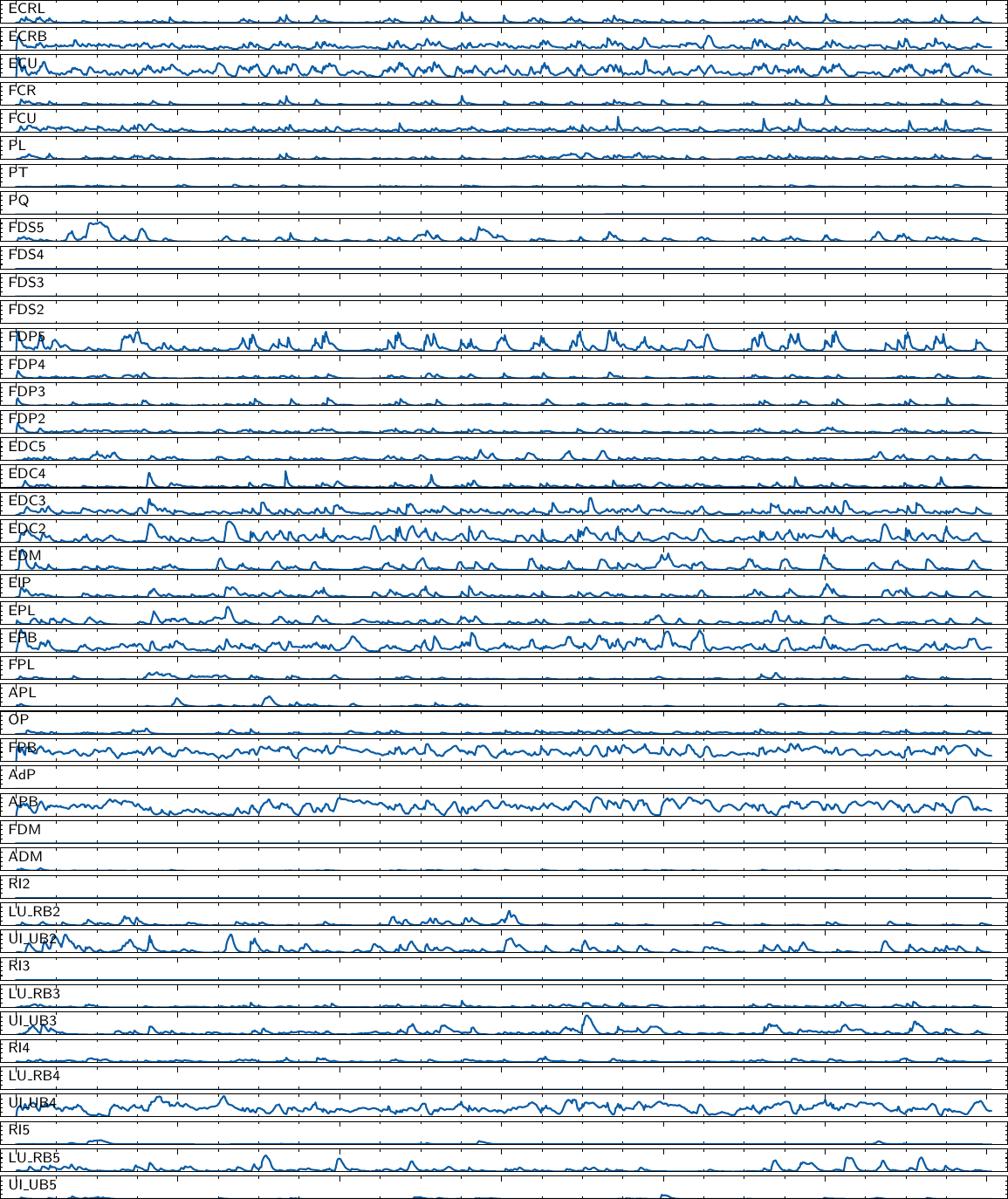}
    \caption{Muscle activations generated for left hand control during the policy playing \textit{F\"ur Elise}.}
    \label{fig:act}
\end{figure*}

\begin{figure*}
    \centering
    \includegraphics[width=0.195\linewidth]{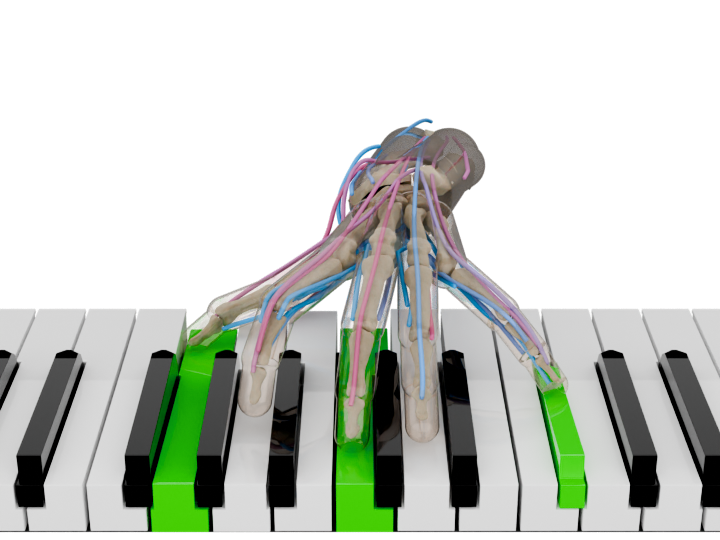}\hfill\includegraphics[width=0.195\linewidth]{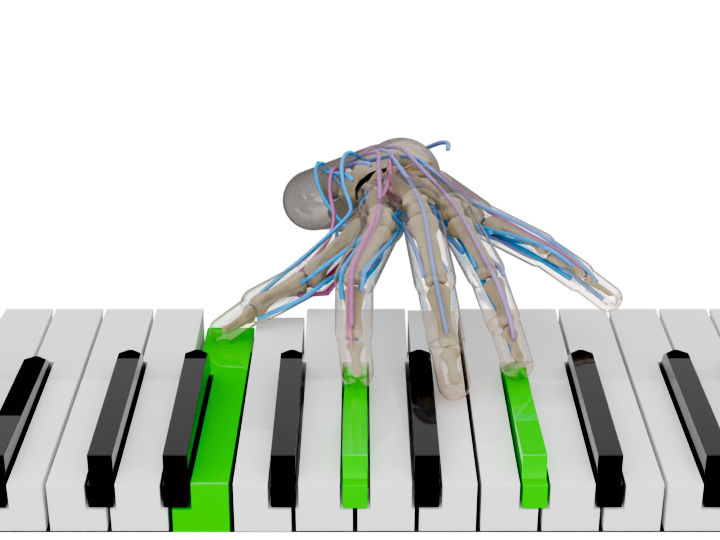}\hfill\includegraphics[width=0.195\linewidth]{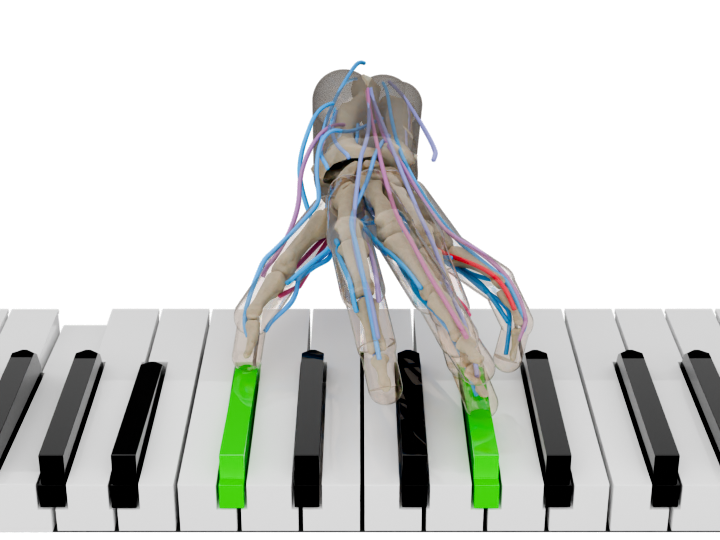}\hfill\includegraphics[width=0.195\linewidth]{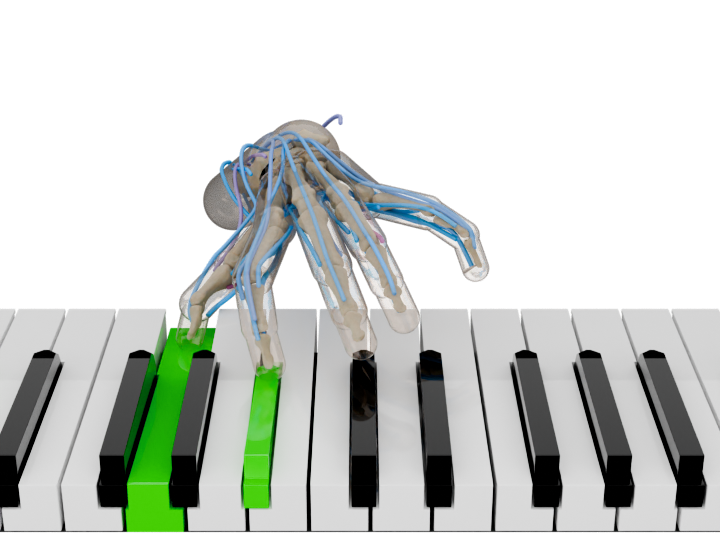}\hfill\includegraphics[width=0.195\linewidth]{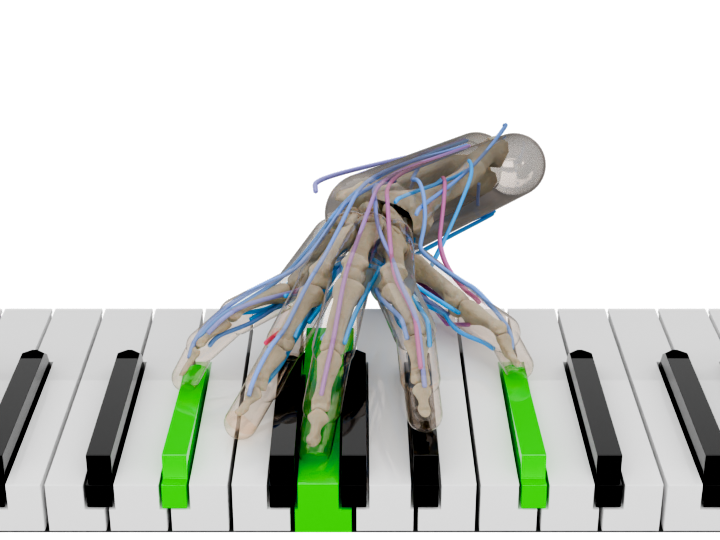}

    \includegraphics[width=0.195\linewidth]{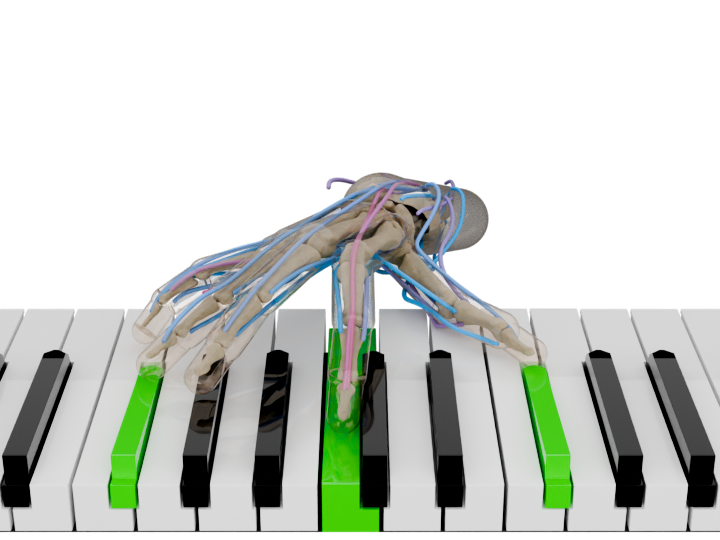}\hfill\includegraphics[width=0.195\linewidth]{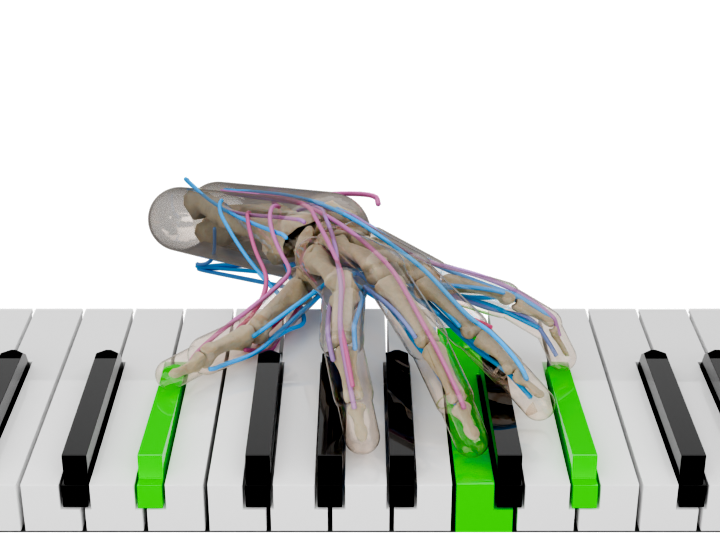}\hfill\includegraphics[width=0.195\linewidth]{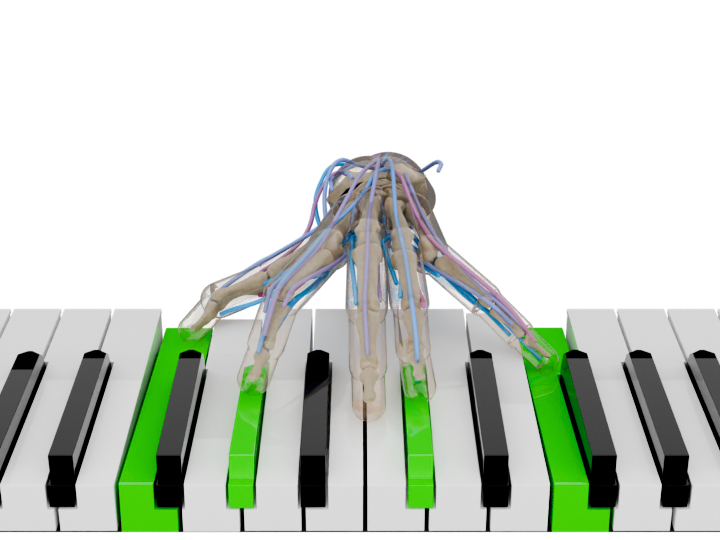}\hfill\includegraphics[width=0.195\linewidth]{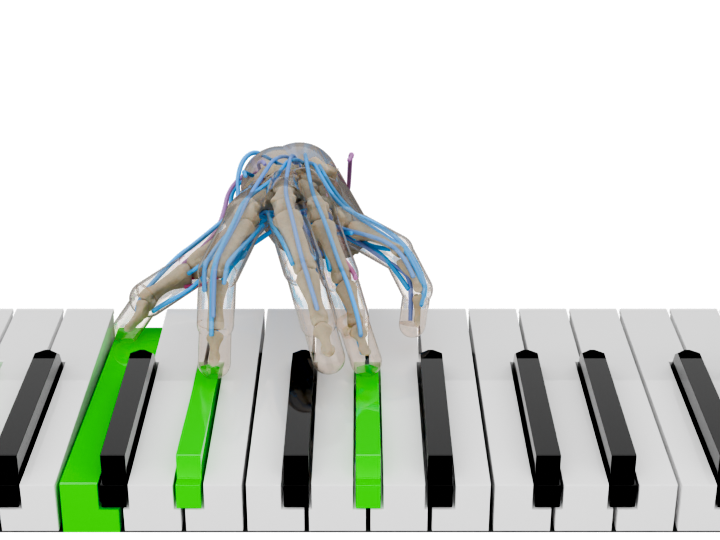}\hfill\includegraphics[width=0.195\linewidth]{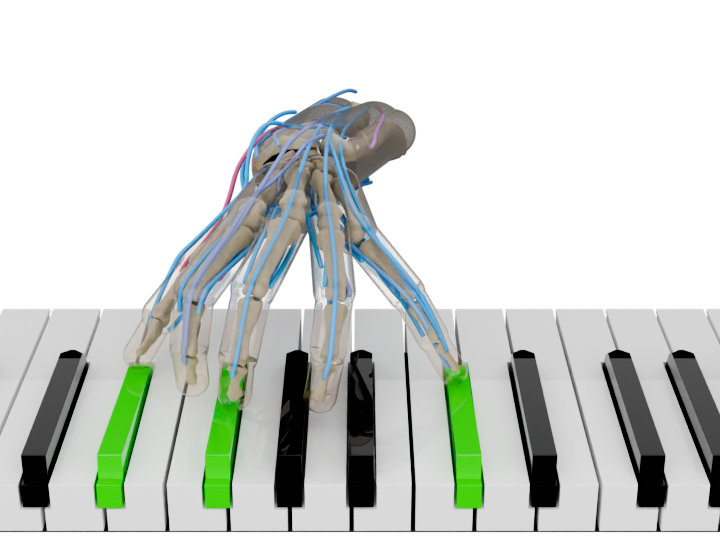}
    \caption{Additional demonstrations of diverse finger poses when multiple keys are pressed at the same time.}
    \label{fig:finger_pose_add}
\end{figure*}

\section{Incorporation of Prior Reference Motion}\label{sec:prior_ref}

We further explore an extension of our approach by introducing a prior reference to facilitate high-level policy learning, as illustrated in Fig.~\ref{fig:net_prior_ref}. 
Correspondingly, in Fig.~\ref{fig:ablation_marl_prior}, we include an additional baseline (MARL+P.Prior) that leverages motions generated by the joint-driven control policy (c.f. Fig.~\ref*{fig:ablation_space}) as the prior reference. 
The prior is incorporated within the same training framework of multi-agent reinforcement learning and serves as an auxiliary guide for exploration. 
As shown in the figure, incorporating this prior accelerates convergence and yields a modest but consistent improvement in final performance. 
This suggests that prior motion references can complement our latent-space control formulation, while the core advantages of decentralized multi-agent learning remain dominant.

\section{Additional Results}

Figure~\ref{fig:latent} visualizes the distribution of VAE latent variables $\mathbf{z}$, with highlights indicating the subsets of latents used for specific music scores. The latent space exhibits a broadly structured organization, where latents corresponding to the same music score tend to form coherent clusters. This behavior suggests that the VAE captures piece-specific motion patterns while maintaining a shared latent structure that supports smooth interpolation and generalization across different musical contexts.

In Fig.~\ref{fig:act}, we show the muscle activations, while the high-level controller plays the tested score of \textit{F\"ur Elise}. 
While abstracted away from detailed muscle states,
the high-level controller can effectively rely on the low-level controller to generate sparse activations, resulting in energy-efficient control, while synthesizing motions to accurately play the piece of music outside the reference data.
We refer to the supplementary video for animated results of the tested repertoire.

\end{document}